\documentclass[pra,twocolumn,aps,amssymb,showpacs,superscriptaddress]{revtex4-1}

\usepackage{epsfig}
\usepackage{graphicx}
\usepackage{amsmath}
\usepackage{amssymb}
\usepackage{amsmath,amssymb}
\usepackage{enumerate}
\usepackage{hyperref}
\usepackage[normalem]{ulem}
\usepackage{cancel}

\usepackage{color}
\definecolor{rosso}{rgb}{1,0,0}
\definecolor{verde}{rgb}{0,1,0}
\definecolor{blue}{rgb}{0,0,1}
\definecolor{verdescuro}{rgb}{0,0.5,0.5}
\definecolor{rossoscuro}{rgb}{0.7,0.3,0}
\definecolor{bluscuro}{rgb}{0.3,0,0.7}
\definecolor{magenta}{rgb}{1,0,1}

\begin{document}

\title{Pair correlations in the normal phase of an attractive Fermi gas}

\author{M. Pini}
\affiliation{School of Science and Technology, Physics Division, Universit\`{a} di Camerino, 62032 Camerino (MC), Italy}
\author{P. Pieri}
\affiliation{School of Science and Technology, Physics Division, Universit\`{a} di Camerino, 62032 Camerino (MC), Italy}
\affiliation{INFN, Sezione di Perugia, 06123 Perugia (PG), Italy}
\author{M. J\"{a}ger}
\affiliation{Institut f\"{u}r Quantenmaterie and Center for Integrated Quantum Science and Technology $(\mathrm{IQ}^{ST})$, Universit\"{a}t Ulm, 89069 Ulm, Germany}
\author{J. Hecker Denschlag}
\email{johannes.denschlag@uni-ulm.de}
\affiliation{Institut f\"{u}r Quantenmaterie and Center for Integrated Quantum Science and Technology $(\mathrm{IQ}^{ST})$, Universit\"{a}t Ulm, 89069 Ulm, Germany}
\author{G. Calvanese Strinati}
\email{giancarlo.strinati@unicam.it}
\affiliation{School of Science and Technology, Physics Division, Universit\`{a} di Camerino, 62032 Camerino (MC), Italy}
\affiliation{INFN, Sezione di Perugia, 06123 Perugia (PG), Italy}
\affiliation{CNR-INO, Istituto Nazionale di Ottica, Sede di Firenze, 50125 (FI), Italy}


\begin{abstract}
In a recent paper [Phys. Rev. A {\bf 99}, 053617 (2019)], the total number of fermion pairs in a spin-balanced two-component Fermi gas of $^{6}$Li atoms was experimentally probed in the normal phase above the superfluid critical temperature, in order to investigate the sectors of pseudogap and preformed-pair in the temperature-coupling phase diagram. Here, we present a theoretical account of these experimental results in terms of an \emph{ab-initio} self-consistent \mbox{$t$-matrix} calculation, which emphasizes the role of the pair-correlation function between opposite-spin fermions at equilibrium. 
Good agreement is found between the available experimental data and the theoretical results obtained with no adjustable parameter.
\end{abstract}

\maketitle

\section{Introduction} 
\label{sec:introduction}

Preformed pairs are meant to be bound states which form above the critical temperature of a fermionic superfluid  \cite{Micnas-1990,Capone-2016}. 
They are usually associated with the occurrence of a pseudo-gap which can be viewed as a carry-over of the pairing gap in the superfluid phase to the normal phase \cite{Levin-2005}. 
Although in the limit of low density and strong fermionic attraction, a preformed pair can be approximately described by a bound state of two fermions of opposite spin, in general it has 
intrinsically a many-body nature. 
In order to take into account the many-body character of a pair, it is convenient to describe the pair problem in terms of correlations between the fermions. 
These correlations are a non-trivial function of temperature, particle density, and the inter-particle coupling.

Preformed pairs were recently studied in an experiment with a spin-balanced two-component Fermi gas of $^6$Li in the normal phase \cite{Ulm-Cam-2019},  where the number of fermion pairs $N_{\mathrm{p}}$ was determined by converting all atom pairs to tightly-bound diatomic molecules which afterwards were detected. 
The pairing fraction $N_{\mathrm{p}}/N_{\sigma}$ (where $N_{\sigma}$ is the number of all atoms per spin-state) was reported for various temperatures and couplings on the BEC side 
of the BCS-BEC crossover.

A preliminary theoretical account of the pairing fractions was already presented in Ref.~\cite{Ulm-Cam-2019}, which was obtained by a statistical model of non-interacting atoms and molecules 
at equilibrium \cite{Chin-Grimm-2004,Eagles-1969} as well as by an \emph{ab-initio} diagrammatic $t$-matrix approach \cite{PPS-2019}. 
However, the comparison between experiment and theory presented in Ref.~\cite{Ulm-Cam-2019} called for further improvements, because the statistical model could not be confidently 
extended to the crossover region and the $t$-matrix calculation was lacking refinements which turned out to be important for the crossover region.

Here, we present an improved account of the theoretical approach. 
We investigate correlations between spin-up and spin-down fermions at thermal equilibrium. 
On the basis of this, we derive a meaningful definition and measure for preformed pairs. 
We calculate thermodynamic quantities such as the pairing fraction, rather than dynamical quantities such as the pseudo-gap. 
Nevertheless, the pseudo-gap physics is well contained in our approach. 
As a consequence, the results of our quantum many-body approach in the crossover region differ significantly from the ones of the statistical atom-molecule model where the fermionic character of the pairs is neglected. 
In general, we find good agreement between theory and experiment, giving us confidence on the validity of our approach.

Our detailed theoretical interpretation of the experimental data of Ref.~\cite{Ulm-Cam-2019} and new insights on the separation between the molecular and pseudo-gap regimes are 
the main results of this paper. 
In addition, we calculate for a homogeneous Fermi gas (i) the pair correlation function, (ii) Tan's contact (a quantity that sets the overall scale of the pair correlation function), and (iii) the pairing fraction. These three quantities are calculated for different temperatures and couplings across the BCS-BEC crossover. 
For a trapped system, we also report density profiles and compare them to experimental measurements, and we provide the superfluid critical temperature across the BCS-BEC crossover.

It should be mentioned that the temperature dependence of the contact in the homogeneous case and of the density profiles in the trapped case were already reported in Refs.~\cite{Enss-2011} and \cite{Haussmann-2008} within the same self-consistent $t$-matrix approach of our work, albeit only for the unitary case.  
We have verified that for this case our results fully agree with the published ones. 

The paper is organized as follows.
Section \ref{sec:theoretical_approach} describes the theoretical approach. 
Section \ref{sec:numerical_results_homogeneous} presents calculated pair fractions $N_{\mathrm{p}}/N_{\sigma}$ for the homogeneous system.
Section \ref{sec:comparison_with_experiment} compares these results to the experimental data of Ref.~\cite{Ulm-Cam-2019} after suitable averaging for the trap.
Section \ref{sec:conclusions} presents our conclusions.
Appendix \ref{sec:appendix-conserving} discusses the use of conserving approximations for the many-body structure of the pair fraction.
Appendix \ref{sec:appendix-model} highlights the circumstances under which the many-body approach to the pair fraction reduces to that of the statistical model.
Finally, Appendix \ref{sec:appendix-scattering-length} obtains the critical temperature of a trapped low-density Bose gas.
Throughout the paper, we set $\hbar =1$.

\section{Theoretical approach} 
\label{sec:theoretical_approach}

The theoretical approach that we set up to account for the experimental results of Ref.~\cite{Ulm-Cam-2019} on the pair fraction builds on the following ingredients:
(i) The definition of the many-body propagator for composite bosons introduced in Appendix A of Ref.~\cite{Andrenacci-2003};
(ii) The formalism developed in Ref.~\cite{Palestini-2014} to calculate the pair correlation function of opposite-spin fermions also in the normal phase;
(iii) The experience recently nurtured in Ref.~\cite{PPS-2019} on the fully self-consistent solution of the $t$-matrix approach to a Fermi gas with an attractive inter-particle interaction.

This Fermi gas is made to span the BCS-BEC crossover by varying the (dimensionless) coupling parameter $(k_{F} a_{F})^{-1}$, where $k_{F}=(3 \pi^{2} n)^{1/3}$ is the Fermi wave vector associated with the number density $n$ and $a_{F}$ is the scattering length of the two-fermion problem \cite{Physics-Reports-2018}.
In practice, the crossover between the BCS and BEC regimes is exhausted within the range $-1 \lesssim (k_{F} a_{F})^{-1} \lesssim +1$ about unitarity where $(k_{F}a_{F})^{-1} = 0$.
In the following, we shall mostly be interested in the coupling region $0 \lesssim (k_{F} a_{F})^{-1} \lesssim +1.5$ on the BEC side of unitarity for which the experimental data of Ref.~\cite{Ulm-Cam-2019} are available.

\vspace{0.05cm}
\begin{center}
{\bf A. Outline of the theoretical expressions to be related with the experimental data}
\end{center}

Strictly speaking, a pair of spin-up and spin-down fermions can be regarded as a purely bosonic entity only in the BEC regime and at sufficiently low temperatures. In all other cases, one should search for \emph{correlations} between fermions and define the occurrence of pairing accordingly. 
Adopting this point of view, which applies also to the so-called Cooper pairs in the BCS regime, is definitely required on the BEC side of unitarity in the normal phase, where the experimental data reported in Ref.~\cite{Ulm-Cam-2019} were collected.
To this end, a suitable definition is needed of what would loosely speaking be referred to as a ``preformed pair'' in the normal phase of a fermionic superfluid.
This definition should be based on a quantum many-body approach where fermions are the elementary constituents of the theory, with no \emph{a priori} reference to the preformed pairs themselves.

We begin by introducing the bosonic propagator
\begin{equation}
\mathcal{G}_{B}(x,x')  =  - \langle T_{\tau}[\Psi_{B}(x) \Psi^{\dagger}_{B}(x')] \rangle \, ,
\label{defin-G}
\end{equation}
where $x=(\mathbf{r},\tau)$ groups the spatial position $\mathbf{r}$ and imaginary time $\tau$, $\Psi_{B}(\mathbf{r})$ is a bosonic field operator, $T_{\tau}$ the time-ordered operator, and
$\langle \cdots \rangle$ a thermal average taken at temperature $T$ \cite{FW-1971}.
In terms of this propagator, the total number of bosons is given by
\begin{eqnarray}
N_{\mathrm{p}} & = & - \int \! d\mathbf{r} \, \mathcal{G}_{B}(x,x^{+})
\nonumber \\
& = & - \int \! d\mathbf{r} \! \int \! \frac{d\mathbf{q}}{(2 \pi)^{3}} \frac{1}{\beta} \sum_{\nu} e^{i \Omega_{\nu} \eta} \, \mathcal{G}_{B}(\mathbf{q},\Omega_{\nu}) \, ,
\label{bosonic-total-density}
\end{eqnarray}
\noindent
where $\mathbf{q}$ is a wave vector, $\Omega_{\nu} = 2 \pi \nu / \beta$ ($\nu$ integer) a bosonic Matsubara frequency with $\beta = (k_{B} T)^{-1}$ and $k_{B}$ the
Boltzmann constant, and $\eta = 0^{+}$.
In the last line of Eq.~(\ref{bosonic-total-density}) a homogeneous system has been assumed, for which one may simply write $N_{\mathrm{p}} = \mathcal{V} \, n_{\mathrm{p}}$ where
$\mathcal{V}$ is the volume occupied by the system and $n_{\mathrm{p}}$ the boson density.

To the extent that the bosonic entities we are considering are made up of fermion pairs, the bosonic operator $\Psi_{B}({\mathbf r})$ has to be related to its fermionic counterparts
$\psi_{\sigma}({\mathbf r})$, where $\sigma = (\uparrow,\downarrow)$ is the spin projection.
This can be achieved by setting
\begin{equation}
\Psi_{B}({\mathbf r}) = \int \! d\boldsymbol{\rho} \, \phi(\boldsymbol{\rho}) \, \psi_{\downarrow}({\mathbf r}-\boldsymbol{\rho}/2) \, \psi_{\uparrow}({\mathbf r}+\boldsymbol{\rho}/2)
\label{definition-boson-fermion}
\end{equation}
where $\phi(\boldsymbol{\rho})$ is a suitable function that should itself embody the correlations within a fermion pair we are after.

On physical grounds, at sufficiently low temperature in the BEC regime it is reasonable to take $\phi(\boldsymbol{\rho})$ as the (normalized) bound-state wave function of the fermionic two-body problem
\emph{in vacuum}, namely,
\begin{equation}
\phi(\boldsymbol{\rho}) = \frac{1}{\sqrt{2 \pi a_{F}}}  \frac{e^{-\rho/a_{F}}}{\rho}
\label{psi-rho-BEC}
\end{equation}
where $\rho = |\boldsymbol{\rho}|$, whose Fourier transform reads
\begin{equation}
\phi({\mathbf p})  =  \sqrt{\frac{8 \pi}{a_{F}}} \frac{1}{{\mathbf p}^{2}  +  a_{F}^{-2}}  \, .
\label{psi-k-BEC}
\end{equation}
As already mentioned, the definition (\ref{definition-boson-fermion}) together with the expression (\ref{psi-rho-BEC}) was originally used in Ref.~\cite{Andrenacci-2003} to describe condensed composite bosons well below the superfluid transition temperature $T_{c}$ with fermions treated within the mean-field approximation \cite{BCS-1957}.
The same combination of the expressions (\ref{definition-boson-fermion}) and (\ref{psi-rho-BEC}) was then utilized in Ref.~\cite{Ulm-Cam-2019}, aiming to account for the quantity $N_{\mathrm{p}}$ 
of Eq.~(\ref{bosonic-total-density}) on the BEC side of unitarity in the normal phase above $T_{c}$, even up to a few times the Fermi temperature $T_{F}$.
In addition, in this case fermions were treated within the self-consistent $t$-matrix approach \cite{PPS-2019}, with a further trap averaging to comply with the experimental procedure of 
Ref.~\cite{Ulm-Cam-2019}.

To account for the experimental data of Ref.~\cite{Ulm-Cam-2019} in a comprehensive way, however, the function $\phi(\boldsymbol{\rho})$ with which the projection is performed
in Eq.~(\ref{definition-boson-fermion}) should acquire a more general form than the expression (\ref{psi-rho-BEC}), which is expected to be valid only in the BEC regime at low temperature.
Accordingly, in what follows (cf.~Section \ref{sec:numerical_results_homogeneous}-A) we will replace the expression (\ref{psi-rho-BEC}) by a more general form obtained from the pair correlation function studied in Ref.~\cite{Palestini-2014}, a form which can thus be utilized even past unitarity towards the BCS regime and up to a temperature of even several times $T_{F}$.

In addition, we shall see below 
(cf.~Section \ref{sec:theoretical_approach}-B) that in the diagrammatic expansion of the expressions (\ref{defin-G}) and (\ref{definition-boson-fermion}) one should also retain an ``unbound'' term that was disregarded in the analysis of Ref.~\cite{Ulm-Cam-2019} since it is negligible in the BEC limit.

It turns out (cf.~Section \ref{sec:comparison_with_experiment}-B) that both these refinements (namely, the inclusion of the above unbound term and the improvement of the expression (\ref{psi-rho-BEC}) in terms of the pair correlation function) improve the comparison with the experimental data of Ref.~\cite{Ulm-Cam-2019}, especially just on the BEC side of unitarity.
This comparison will also make it possible to distinguish between the pseudo-gap and the molecular regimes mentioned in the Introduction. Specifically, we argue that the molecular regime should be reached when the unbound term contributes in a negligible way to the quantity $N_{\mathrm{p}}$ of Eq.~(\ref{bosonic-total-density}).
 
\vspace{0.05cm}
\begin{center}
{\bf B. Diagrammatic approach to the pair fraction}
\end{center}

We pass now to describe the diagrammatic approach that we have adopted for the calculation of the expressions (\ref{defin-G})-(\ref{definition-boson-fermion}).
Although we are interested in the normal phase above $T_{c}$ which the experimental data of Ref.~\cite{Ulm-Cam-2019} are restricted to, we find it convenient to adopt the Nambu representation of the fermionic field operators
\begin{equation} 
\Psi({\mathbf r})  =  \left( \begin{array}{c}
                              \psi_{\uparrow}({\mathbf r}) \\
                              \psi_{\downarrow}^{\dagger}({\mathbf r}) 
                              \end{array} \right) \, ,
\label{Nambu-representation}
\end{equation}
in terms of which the diagrammatic approach for the superfluid phase below $T_{c}$ is usually formulated \cite{Schrieffer-1964}.
This is mainly because the concept of fermion pairing originates from the superfluid phase \cite{BCS-1957}, from which it can be extrapolated to the normal phase in the context of the BCS-BEC crossover \cite{Physics-Reports-2018} under suitable circumstances, like in the present case.
In addition, through the Nambu representation (\ref{Nambu-representation}) one finds it easier to deal with the issue of conserving approximations for a fermionic superfluid \cite{Baym-1962}. 
This proves important when one selects the set of diagrams that would describe at best the physical problem of interest, with the condition that their numerical implementation remains affordable.
We shall discuss this issue in Appendix \ref{sec:appendix-conserving}.

In terms of the Nambu representation (\ref{Nambu-representation}), one writes for the fermionic single-particle Green's function
\begin{equation}
{\mathcal G}(1,2)  =  -  \langle T_{\tau}[\Psi(1) \Psi^{\dagger}(2)] \rangle
\label{G-1}
\end{equation}
and for the fermionic two-particle Green's function
\begin{equation}
{\mathcal G}_{2}(1,2;1',2')  =  \langle T_{\tau}[\Psi(1) \Psi(2) \Psi^{\dagger}(2') \Psi^{\dagger}(1')] \rangle \, ,
\label{G-2}
\end{equation}
with the short-hand notation $1=({\mathbf r}_{1}, \tau_{1}, \ell_{1})$ and so on, where the Nambu index $\ell=(1,2)$ refers to the upper or lower component in the expression (\ref{Nambu-representation}). 
Here, ${\mathcal G}_{2}$ is related to the two-particle correlation function
\begin{equation}
L(1,2;1',2')  = {\mathcal G}_{2}(1,2;1',2') - {\mathcal G}(1,1') \, {\mathcal G}(2,2')
\label{L}
\end{equation}
which satisfies the Bethe-Salpeter equation \cite{Andrenacci-2003,Baym-1962,Strinati-RNC}
\begin{eqnarray}
L(1,2;1',2') & = & - {\mathcal G}(1,2') {\mathcal G}(2,1') +  \int \!d3456 \; {\mathcal G}(1,3)
\nonumber   \\
& \times & {\mathcal G}(6,1')  \Xi(3,5;6,4) L(4,2;5,2')
\label{Bethe-Salpeter}
\end{eqnarray}
where
\begin{equation}
\Xi(1,2;1',2')  =  \frac{\delta \Sigma(1,1')}{\delta {\mathcal G}(2',2)}
\label{2-p-int}
\end{equation}
is an effective two-particle interaction with $\Sigma$ the fermionic self-energy.
Equation (\ref{Bethe-Salpeter}) can be formally solved in terms of the many-particle T-matrix, defined as the solution to the equation \cite{Andrenacci-2003,Baym-1962,Strinati-RNC}
\begin{eqnarray}
T(1,2;1',2') & = & \Xi(1,2;1',2') + \int \!d3456 \; \Xi(1,4;1',3)
\nonumber   \\
& \times & {\mathcal G}(3,6) {\mathcal G}(5,4)  T(6,2;5,2') \, ,
\label{many-p-T-matrix}
\end{eqnarray}
by writing
\begin{eqnarray}
& -  & L(1,2;1',2') = {\mathcal G}(1,2')  {\mathcal G}(2,1')
\label{L-T} \\
& + & \int d3456 \; {\mathcal G}(1,3) {\mathcal G}(6,1')  T(3,5;6,4) {\mathcal G}(4,2')  {\mathcal G}(2,5) .
\nonumber 
\end{eqnarray}

The above equations hold quite generally, regardless of the specific approximation for the kernel $\Xi$ defined in Eq.~(\ref{2-p-int}).
In particular, to the BCS approximation $\Sigma_{\mathrm{BCS}}$ for the self-energy there corresponds the kernel:
\begin{eqnarray}
\Xi_{\mathrm{BCS}}(1,2;1',2') & = & \frac{\delta \Sigma_{\mathrm{BCS}}(1,1')}{\delta {\mathcal G}_{\mathrm{BCS}}(2',2)}
\label{csibcs} \\
& = & -  \tau^{3}_{\ell_{1}\ell_{2'}} \delta(x_{1}-x_{2'}) v(x_{1}^{+}-x_{1'})
\nonumber \\
& \times & \delta(x_{1'}-x_{2})  \tau^{3}_{\ell_{1'}\ell_{2}}  (1  - \delta_{\ell_{1}\ell_{1'}})
\nonumber
\end{eqnarray}
where only the off-diagonal terms of the BCS self-energy have been retained following a common practice.
In the expression (\ref{csibcs}), $\tau^{3}$ is the third Pauli matrix \cite{Schrieffer-1964}, $x_{1}=({\mathbf r}_{1}, \tau_{1})$ and so on, and $v(x_{1}^{+}-x_{1'}) = \delta(\tau_{1}^{+}-\tau_{1'}) v({\mathbf r_{1}}-{\mathbf r_{1'}})$ is the attractive fermionic interaction.
For the ultra-cold Fermi atoms of interest, one takes $v({\mathbf r_{1}}-{\mathbf r_{1'}}) = v_{0} \, \delta({\mathbf r_{1}}-{\mathbf r_{1'}})$ of the contact form, where the (negative) strength $v_{0} $ is further eliminated in favor of the scattering length $a_{F}$ through a standard regularization procedure \cite{Physics-Reports-2018}.

We return at this point to the expression (\ref{defin-G}) of the bosonic propagator $\mathcal{G}_{B}$ with the definition (\ref{definition-boson-fermion}) for the bosonic field, which we rewrite in the Nambu representation (\ref{Nambu-representation}).
The following compact form then results for $\mathcal{G}_{B}$ in terms of the two-particle correlation function (\ref{L}):
\begin{equation}
\mathcal{G}_{B}({\mathbf r} \tau,{\mathbf r}' \tau')  =  -  \int \!\! d\boldsymbol{\rho} \!\! \int \!\! d\boldsymbol{\rho}'  \phi(\boldsymbol{\rho}) \phi^{*}(\boldsymbol{\rho}') \, L(1,2;1',2')            
\label{G-L}
\end{equation}
with the identification $1=({\mathbf r}+\boldsymbol{\rho}/2,\tau,\ell=1)$, $2=({\mathbf r}'-\boldsymbol{\rho}'/2,\tau',\ell=2)$, $1'=({\mathbf r}-\boldsymbol{\rho}/2,\tau^{+},\ell=2)$,
and $2'=({\mathbf r}'+\boldsymbol{\rho}'/2,\tau'^{+},\ell=1)$.
Hereafter, it will be understood that only the terms that survive once carried over from below to above $T_{c}$ will be retained in the expression (\ref{G-L}).
Accordingly, in passing from Eq.(\ref{L}) to Eq.(\ref{G-L}) we have neglected the second term on the right-hand side of Eq.(\ref{L}), which corresponds to the (square magnitude of the) condensate amplitude
and thus vanishes above $T_{c}$ \cite{Andrenacci-2003}.

In addition, it will be shown in Appendix \ref{sec:appendix-conserving} that, due to the specific identification of the Nambu indices relevant to Eq.~(\ref{G-L}), the many-particle T-matrix of Eq.~(\ref{many-p-T-matrix}) 
which solves the Bethe-Salpeter equation for $L$ can be built \emph{only} in terms of the effective two-particle interaction $\Xi$ of the form (\ref{csibcs}) \cite{footnote-GMB}.
This leaves us with the freedom of endowing the fermionic single-particle Green's function $\mathcal{G}$ of Eq.~(\ref{G-1}) with a suitable \emph{additional} self-energy $\Sigma$ to be selected on physical grounds, without being forced to introduce at the same time related additional terms in the kernel $\Xi$ via Eq.(\ref{2-p-int}).

With these considerations in mind, we have selected this additional self-energy of the form of the fully self-consistent $t$-matrix approach, whose performance in the normal phase above $T_{c}$ 
has been recently tested against those of the non-self-consistent as well as of other partially self-consistent $t$-matrix approaches \cite{PPS-2019}, with the result that the fully self-consistent one performs best at least as far as thermodynamic quantities are concerned.
To the extent that the quantity $N_{\mathrm{p}}$ given by the expression (\ref{bosonic-total-density}) of interest here is itself a thermodynamic quantity (consistently with the fact that no analytic continuation from Matsubara  to real frequencies is required to calculate it), this choice for $\Sigma$ within the fully self-consistent $t$-matrix approach appears to be adequate for our purposes.
In addition, the BCS self-energy $\Sigma_{\mathrm{BCS}}$, which has served to obtain the kernel $\Xi_{\mathrm{BCS}}$ of Eq.~(\ref{csibcs}), vanishes identically in the normal phase and no longer 
needs to be considered in what follows.

For a homogeneous system, we can further make use of the Fourier representation and rewrite Eq.(\ref{G-L}) as:
\begin{eqnarray}
& & \mathcal{G}_{B}(\mathbf{q},\Omega_{\nu}) = -  \int \! \frac{d{\mathbf p}}{(2\pi)^{3}} \frac{1}{\beta} \sum_{n}  \int \! \frac{d{\mathbf p}'}{(2\pi)^{3}}  \frac{1}{\beta} \sum_{n'} 
\nonumber \\
& \times & \phi({\mathbf p}+{\mathbf q}/2) \phi({\mathbf p}'+{\mathbf q}/2) \, L^{11}_{22} ({\mathbf p}\omega_{n},{\mathbf p}'\omega_{n'};\mathbf{q}\Omega_{\nu})
\label{G_B-FT}  
\end{eqnarray}
where $\omega_{n} = (2n+1)\pi/\beta$ ($n$ integer) is a fermionic Matsubara frequency (the conventions for the Nambu indices are specified in Fig.~\ref{Figure-10} of
Appendix \ref{sec:appendix-conserving}).
The expression (\ref{G_B-FT}) will be utilized in Eq.~(\ref{bosonic-total-density}) to obtain the number of pairs $N_{\mathrm{p}}$.
Solving then for the many-particle T-matrix of Eq.~(\ref{many-p-T-matrix}) as described above and entering the result in Eq.~(\ref{L-T}) for $L$, Eq.~(\ref{G_B-FT}) reduces eventually to the form:
\begin{equation}
\mathcal{G}_{B}(\mathbf{q},\Omega_{\nu})  =  - \mathcal{F}_{2}(\mathbf{q},\Omega_{\nu}) - \mathcal{F}_{1}(\mathbf{q},\Omega_{\nu})^{2} \, \Gamma(\mathbf{q},\Omega_{\nu}) \, .
\label{G_B-FT-sc}
\end{equation}
Here, 
\begin{eqnarray}
\mathcal{F}_{j}(\mathbf{q},\Omega_{\nu}) & = & \int \! \frac{d{\mathbf p}}{(2\pi)^{3}} \, \phi({\mathbf p}+{\mathbf q}/2)^{j}   
\label{def-F-j} \\
& \times & \frac{1}{\beta} \sum_{n} \mathcal{G}(\mathbf{p}+\mathbf{q},\omega_{n}+\Omega_{\nu}) \mathcal{G}(-\mathbf{p},-\omega_{n})     
\nonumber
\end{eqnarray}
are ``form factors'' associated with the particle-particle bubble where $j=(1,2)$, and
\begin{equation}
\Gamma(\mathbf{q},\Omega_{\nu}) = - \left( \frac{m}{4 \pi a_{F}} + R_{\mathrm{pp}}(\mathbf{q},\Omega_{\nu}) \right)^{-1}
\label{Gamma}
\end{equation}
is the particle-particle propagator in the normal phase where
\begin{eqnarray}
R_{\mathrm{pp}}(\mathbf{q},\Omega_{\nu}) & = & \int \! \frac{d{\mathbf p}}{(2\pi)^{3}} \frac{1}{\beta} \sum_{n}  \mathcal{G}(\mathbf{p}+\mathbf{q},\omega_{n}+\Omega_{\nu}) \mathcal{G}(-\mathbf{p},-\omega_{n}) 
\nonumber \\
& - & \int \! \frac{d{\mathbf p}}{(2\pi)^{3}} \frac{m}{\mathbf{p}^{2}} 
\label{pp-bubble} 
\end{eqnarray}
is the regularized particle-particle bubble \cite{Physics-Reports-2018}.
We emphasize again that the fermionic single-particle Green's functions $\mathcal{G}$ entering the expressions (\ref{def-F-j}) and (\ref{pp-bubble}) are meant to be obtained within the self-consistent $t$-matrix 
approach in the normal phase \cite{PPS-2019}.

What is still left to be specified is the form of the wave function $\phi(\mathbf{p})$ that enters Eq.~(\ref{def-F-j}).
We have already mentioned that, in the theoretical diagrammatic approach to $N_{\mathrm{p}}$ presented in Ref.~\cite{Ulm-Cam-2019}, $\phi(\mathbf{p})$ was taken of the form (\ref{psi-k-BEC})
corresponding to the fermionic two-body problem.
With this choice, however, meaningful results could be obtained only towards the BEC edge of the BEC side of the unitary region.
To overcome this limitation, here we adopt a more general form for $\phi(\mathbf{p})$ which will be obtained from the pair correlation function, as discussed in Section \ref{sec:numerical_results_homogeneous}-A below.

In addition, in Ref.~\cite{Ulm-Cam-2019} the first term on the right-hand side of Eq.~(\ref{G_B-FT-sc}) was not retained.
As anticipated in Section \ref{sec:theoretical_approach}-A, this term will be referred to as the ``unbound'' term as opposed to the ``bound'' term discussed below.
Here, we are going to keep this ``unbound'' term and show that it gives a non-negligible contribution to $N_{\mathrm{p}}$, through the pairing correlations contained both in the fermionic single-particle Green's function 
$\mathcal{G}$ and in the wave function $\phi(\mathbf{p})$ that enter the expression (\ref{def-F-j}) with $j=2$.
Accordingly, through this term spin-$\uparrow$ and spin-$\downarrow$ fermions correlate with each other \emph{indirectly} via their separate interaction with the environment.

In contrast, the second term on the right-hand side of Eq.~(\ref{G_B-FT-sc}) is referred to as the ``bound'' term, because in this case spin-$\uparrow$ and spin-$\downarrow$ fermions correlate with each other \emph{directly} through their inter-particle attractive interaction.
The result for $N_{\mathrm{p}}$ obtained from this term will be shown to reduce to that of the statistical model of atom-molecule equilibrium introduced in Refs.~\cite{Chin-Grimm-2004,Eagles-1969}, past the BEC side of the unitary region and for not too high temperatures above $T_{c}$.
The reasons for the success of the statistical atom-molecule model in this sector of the phase diagram will be discussed in Appendix \ref{sec:appendix-model}.

\vspace{0.05cm}
\begin{center}
{\bf C. Single-particle Green's function}
\end{center}

As discussed in Section \ref{sec:theoretical_approach}-B, the single-particle Green's function $\mathcal{G}(\mathbf{p},\omega_{n})$ that enters the expressions (\ref{def-F-j}) and (\ref{pp-bubble})
is taken within the fully self-consistent $t$-matrix approach.
It then reads:
\begin{equation}
\mathcal{G}(\mathbf{p},\omega_{n}) = \left[ \mathcal{G}_{0}(\mathbf{p},\omega_{n})^{-1} - \Sigma(\mathbf{p},\omega_{n}) \right]^{-1} 
\label{self-consistent-G}
\end{equation}
where $\mathcal{G}_{0}(\mathbf{p},\omega_{n}) = [i \omega_{n} - \xi(\mathbf{p})]^{-1}$ is the non-interacting counterpart with $\xi(\mathbf{p}) = \mathbf{p}^{2}/(2m) - \mu$ ($m$ being the fermion mass and $\mu$ the chemical potential) and 
\begin{equation}
\Sigma(\mathbf{p},\omega_{n}) = - \int \! \frac{d{\mathbf q}}{(2\pi)^{3}} \frac{1}{\beta} \sum_{\nu} \Gamma(\mathbf{q},\Omega_{\nu}) \,\mathcal{G}(\mathbf{q}-\mathbf{p},\Omega_{\nu}-\omega_{n}) 
\label{t-matrix-self-energy}
\end{equation}
is the self-energy with $\Gamma(\mathbf{q},\Omega_{\nu})$ given by Eqs.~(\ref{Gamma}) and (\ref{pp-bubble}).
The chemical potential is eventually obtained from the fermionic density $n_{\sigma}$ via the relation
\begin{equation}
n_{\sigma} = \! \int \! \frac{d\mathbf{p}}{(2 \pi)^{3}} \frac{1}{\beta} \sum_{n} e^{i \omega_{n} \eta} \, \mathcal{G}(\mathbf{p},\omega_{n}) 
\label{fermionic-density-equation}
\end{equation}
where $n_{\uparrow} = n_{\downarrow} = n/2$ like in Ref.~\cite{Ulm-Cam-2019}.
The numerical calculation of the expressions (\ref{self-consistent-G})-(\ref{fermionic-density-equation}) will be implemented by taking advantage of the detailed procedures recently reported
in Ref.~\cite{PPS-2019}.

In addition, the strong-coupling (BEC) limit of the expressions (\ref{self-consistent-G})-(\ref{fermionic-density-equation}), together with that of the expressions (\ref{bosonic-total-density}) and (\ref{G_B-FT-sc})-(\ref{pp-bubble}), will be examined in Appendix \ref{sec:appendix-model}, to determine under what circumstances the results for $n_{\mathrm{p}}$ and $n_{\sigma}$ obtained by our diagrammatic quantum many-body theory reduce to those of the statistical model of atom-molecule equilibrium developed in Refs.~\cite{Chin-Grimm-2004,Eagles-1969}.

\section{Results for a homogeneous gas} 
\label{sec:numerical_results_homogeneous}

In this Section, we implement the calculation of the bosonic density $n_{\mathrm{p}}$ obtained from Eqs. (\ref{bosonic-total-density}) and (\ref{G_B-FT-sc}) for a homogeneous gas, as a function of coupling and  temperature.
The information gathered in this way will be used in Section \ref{sec:comparison_with_experiment} when dealing with a trapped gas, by performing a trap average within a local-density approach.
At that point it will be possible to compare the theoretical results with the experimental data of Ref.~\cite{Ulm-Cam-2019}.

The main ingredients of the calculation of $n_{\mathrm{p}}$ are the single-particle Green's function $\mathcal{G}(\mathbf{p},\omega_{n})$ and the wave function $\phi({\mathbf p})$ that enter 
Eqs.~(\ref{def-F-j})-(\ref{pp-bubble}).
The calculation of $\mathcal{G}(\mathbf{p},\omega_{n})$ was already considered in Section \ref{sec:theoretical_approach}-C. It thus remains to consider the calculation of the wave function $\phi({\mathbf p})$, as discussed next.

\vspace{0.05cm}
\begin{center}
{\bf A. Pair correlation function}
\end{center}

Our interpretation of the experimental data of Ref.~\cite{Ulm-Cam-2019} rests on the occurrence of \emph{correlations} between spin-up and spin-down fermions at equilibrium.
The preliminary theoretical account of those experimental data presented in Ref.~\cite{Ulm-Cam-2019} took the wave function $\phi({\mathbf p})$ entering Eq.~(\ref{def-F-j}) of the form
(\ref{psi-k-BEC}) associated with the fermionic two-body problem.
This form, however, proves able to account for the correlations between spin-up and spin-down fermions only in the BEC regime of coupling and at low enough temperature.
As anticipated in Section \ref{sec:theoretical_approach}-B, we now consider a more general form for $\phi({\mathbf p})$ which is obtained from the pair correlation function 
\begin{eqnarray}
g_{\uparrow \downarrow}(\boldsymbol{\rho}) & = &
\left\langle  \psi^{\dagger}_{\uparrow}\left(\frac{\boldsymbol{\rho}}{2}\right) \psi^{\dagger}_{\downarrow}\left(-\frac{\boldsymbol{\rho}}{2}\right) 
                   \psi_{\downarrow}\left(-\frac{\boldsymbol{\rho}}{2}\right) \psi_{\uparrow}\left(\frac{\boldsymbol{\rho}}{2}\right) \right\rangle 
\nonumber \\ 
& - & \left( \frac{n}{2} \right)^{2}  \, .                                                                       
\label{definition-pair-correlation-function}
\end{eqnarray}
This function contains information about correlations between fermions of opposite spins at a distance $\rho = |\boldsymbol{\rho}|$ apart.
This quantity was studied in detail in Ref.~\cite{Palestini-2014} throughout the BCS-BEC crossover, both in the superfluid phase below $T_{c}$ and in the normal phase above $T_{c}$.
Here, we consider the formalism of Ref.~\cite{Palestini-2014} above $T_{c}$ and rephrase it in terms of the fully self-consistent $t$-matrix approach that was summarized in 
Section \ref{sec:theoretical_approach}-C.

Within the fully self-consistent $t$-matrix approach, the expression (\ref{definition-pair-correlation-function}) for $g_{\uparrow \downarrow}(\boldsymbol{\rho})$ can be cast in the form \cite{Palestini-2014}:
\begin{eqnarray}
& & g_{\uparrow \downarrow}(\boldsymbol{\rho}) = \int \! \frac{d{\mathbf q}}{(2\pi)^{3}} \frac{1}{\beta} \sum_{\nu} e^{i \Omega_{\nu} \eta} \, \Gamma(\mathbf{q},\Omega_{\nu}) 
\label{pair-correlation-function-above-Tc} \\
& \times & \!\! \int \! \frac{d\mathbf{p}}{(2 \pi)^{3}} \, e^{i \mathbf{p} \cdot \boldsymbol{\rho}} \, \tilde{\Pi}(\mathbf{p};\mathbf{q},\Omega_{\nu}) \, 
                       \int \! \frac{d\mathbf{p}'}{(2 \pi)^{3}} \, e^{-i \mathbf{p}' \cdot \boldsymbol{\rho}} \, \tilde{\Pi}(\mathbf{p}';\mathbf{q},\Omega_{\nu}) 
\nonumber 
\end{eqnarray}
where
\begin{equation}
\tilde{\Pi}(\mathbf{p};\mathbf{q},\Omega_{\nu}) = \frac{1}{\beta} \sum_{n} \, \mathcal{G}(\mathbf{p}+\mathbf{q},\omega_{n}+\Omega_{\nu}) \, \mathcal{G}(-\mathbf{p},-\omega_{n}) \, .
\label{Pi}
\end{equation}
Here, the fully self-consistent $\mathcal{G}$'s are considered, while in the original Ref.~\cite{Palestini-2014} non-interacting $\mathcal{G}_{0}$ corresponding to the non-self-consistent approximation were utilized.

It was also shown in Ref.~\cite{Palestini-2014} that $g_{\uparrow \downarrow}(\boldsymbol{\rho})$ given by the expression (\ref{pair-correlation-function-above-Tc}) recovers the short-range behavior 
related to Tan's contact $C$ \cite{Tan-2008_a,Tan-2008_b,Braaten-2012}
\begin{equation}
g_{\uparrow \downarrow}(\boldsymbol{\rho}) \,\,  _{\overrightarrow{(\boldsymbol{\rho}\rightarrow 0)}} \,\, \frac{C}{(4 \pi)^{2}} \, \left( \frac{1}{\rho^{2}} - \frac{2}{\rho \, a_{F}} + \cdots \right) \, ,
\label{contact-BCS-2}
\end{equation}
such that $\lim_{\rho \to 0} \frac{(4 \pi)^{2}}{C} \rho^{2} \, g_{\uparrow \downarrow}(\rho)= 1$ irrespective of coupling and temperature.
We have reproduced here these analytic results within our fully self-consistent $t$-matrix approach, with the numerical values of $C$ obtained in agreement with Ref.~\cite{PPS-2019}. 

\begin{figure}[t]
\begin{center}
\includegraphics[width=9.0cm,angle=0]{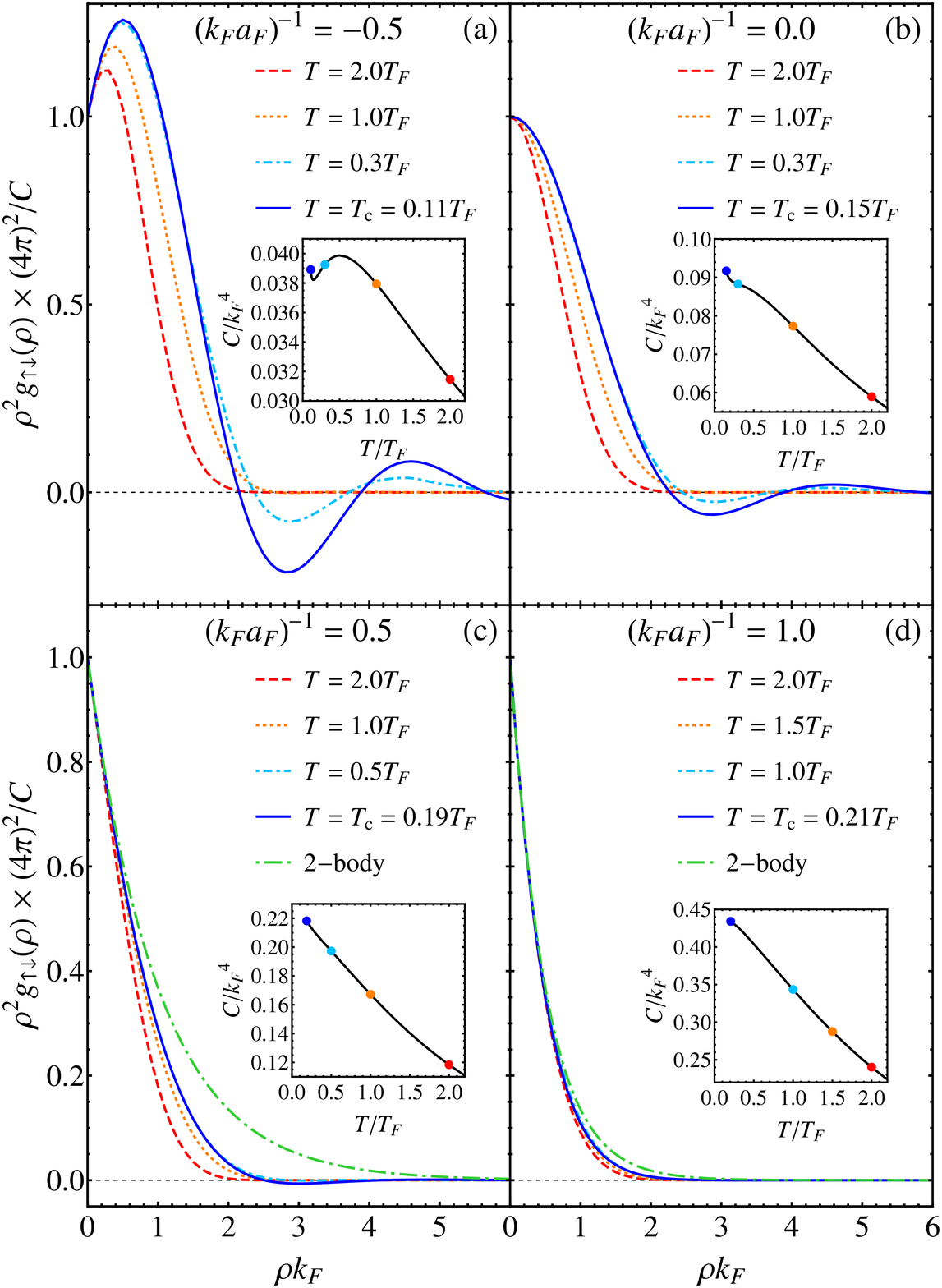}
\caption{(Color online) Spatial profiles of $\rho^{2} \, g_{\uparrow \downarrow}(\rho)$ are shown vs $\rho$ (in units of $k_{F}^{-1}$), for several couplings about unitarity and different temperatures 
                                     in the normal phase. In each panel, the inset gives the dependence of the contact $C$ over an extended range of temperature (in units of the Fermi temperature $T_{F}$), 
                                     where the dots correspond to the temperatures reported in the same panel. In panels (c) and (d), the expression $(2\pi a_F)\rho^{2} |\phi(\rho)|^2=e^{-2\rho/a_F}$ corresponding to the two-body bound state  (\ref{psi-rho-BEC}) is reported for comparison (long dash-dotted lines).}
\label{Figure-1}
\end{center}
\end{figure}

Examples of the spatial profiles of the pair correlation function $g_{\uparrow \downarrow}(\rho)$ are shown in Fig.~\ref{Figure-1}, for several couplings and temperatures above $T_{c}$.
Reported in each inset are also the respective values of the contact $C$, from which the numerical values of $g_{\uparrow \downarrow}(\rho)$ can be explicitly reconstructed.
Note the oscillatory behavior of $g_{\uparrow \downarrow}(\rho)$, which is present on the BCS side at low temperatures but quickly fades away either by moving towards the BEC side or by increasing temperature.
Due to this oscillatory behavior, $g_{\uparrow \downarrow}(\rho)$ may acquire negative values which correspond to a weaker correlation with respect to the uncorrelated value $n_{\uparrow} \, n_{\downarrow} =(n/2)^{2}$ \cite{Palestini-2014}.
This behavior, however, will not affect our argument below, whereby the oscillations about zero (whenever present) will be averaged out.

It can be further verified from the expression (\ref{pair-correlation-function-above-Tc}) that, in the BEC limit and at sufficiently low temperatures, $g_{\uparrow \downarrow}(\rho)$ reduces to the product  
of the density $n_{\sigma} = n/2$ of a single fermionic species times the \emph{square} of the wave function (\ref{psi-rho-BEC}) corresponding to the fermionic two-body problem.
This suggests that the function $\phi(\mathbf{p})$, to be utilized in the form factors (\ref{def-F-j}), can be extracted from  the pair correlation function $g_{\uparrow \downarrow}(\rho)$ also away from the BEC limit and at high temperatures.
To this end, we adopt the following strategy.

We begin by fitting the spatial profiles of the function $\frac{(4 \pi)^{2}}{C} \rho^{2}  g_{\uparrow \downarrow}(\rho)$ of Fig.~\ref{Figure-1} with the expression
\begin{equation}
\rho^{2} \phi(\rho)^{2} = \exp(-2 \rho/a_{F}) \, \exp(- 2 b \rho^{2}) \, ,
\label{square-fitting-function}
\end{equation}
where $b$ is a parameter that depends on coupling and temperature (note that the function (\ref{square-fitting-function}), too, has unit value at $\rho = 0$).
We then take the square root of the expression (\ref{square-fitting-function}) to extract $\phi(\rho)$, and multiply the result by a suitable normalization factor $\mathcal{N}$, thus writing:
\begin{equation}
\phi(\rho) = \mathcal{N}\!\left(a_{F},b \right) \, \frac{ e^{-\rho/a_{F}} }{ \rho } \, \exp(- b \rho^{2})
\label{fitting-function-normalized}
\end{equation}
with
\begin{equation}
\mathcal{N}\!\left(a_{F},b \right) = \frac{1}{\pi^{3/4}} \Big(\frac{b}{2}\Big)^{1/4} \frac{\exp [-\frac{1}{4ba_{F}^{2}}]} {\sqrt{\text{erfc}[ \frac{1}{\sqrt{2b} a_{F}} ]}} 
\end{equation}
where $\text{erfc}(z)$ is the complementary error function of (complex) argument $z$ \cite{AS-NBS-1972}.
Note that the two-body wave function (\ref{psi-rho-BEC}) is recovered for $b\to0$. 
Finally, we take the Fourier transform of the expression (\ref{fitting-function-normalized}) and obtain the desired result:
\begin{eqnarray}
\phi(\mathbf{p}) = \frac{2 \pi^{3/2} \mathcal{N}\!\left(a_{F},b \right)}{\sqrt{b} \,\, p} \,\, \text{Im} \bigg\{ && \!\!\!\!\!\!\! \exp \bigg[\frac{ \big(a_{F}^{-1} -i p \big)^2 }{4 b} \bigg]   \bigg.
\nonumber \\
& \times & \bigg. \text{erfc} \bigg( \frac{a_{F}^{-1} -i p}{2 \sqrt{b}} \bigg) \!\! \bigg\}
\label{fitting-function-normalized-FT}
\end{eqnarray}
where $p = |\mathbf{p}|$.
This expression recovers Eq.~(\ref{psi-k-BEC}) in the limit $b \rightarrow 0$.

\begin{figure}[t]
\begin{center}
\includegraphics[width=8.0cm,angle=0]{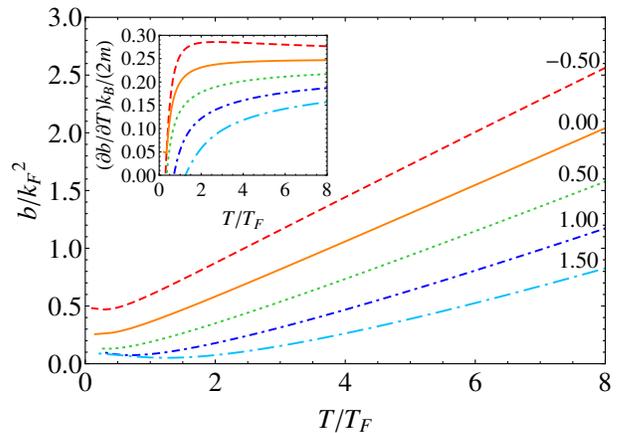}
\caption{(Color online) Temperature dependence of the parameter $b$ of the expressions (\ref{fitting-function-normalized})-(\ref{fitting-function-normalized-FT}) 
                                     for several values (reported above each line) of the coupling $(k_{F}a_{F})^{-1}$ across the BCS-BEC crossover.
                                     The inset shows the derivative of $b$ with respect to $T$ for the same couplings of the main panel to better evidence the high-temperature behavior.}
\label{Figure-2}
\end{center}
\end{figure} 

Figure~\ref{Figure-2} shows the behavior of the parameter $b$ obtained in this way, over a wide range of coupling and temperature relevant to the experiment of Ref.~\cite{Ulm-Cam-2019}.
In particular, for sufficiently high temperature and irrespective of coupling, $b$ is expected to become proportional to $\lambda_{T}^{-2}$ where $\lambda_{T} = \sqrt{\frac{2\pi}{m k_{B} T}}$ is the thermal wavelength.
To evidence this linear behavior of $b$ vs $T$ at high temperature, the inset of Fig.~\ref{Figure-2} plots the derivative of $b$ with respect to $T$ for the same temperature range and couplings of the main panel. 
In all cases, we have found that, at high temperature, this derivative is well reproduced by the expression $\frac{k_{B}}{2m} \frac{\partial b}{\partial T} = 0.25 - 0.175 (k_{F}a_{F})^{-1} \sqrt{T_{F}/T}$.

The fitting function $\phi(\rho)$ given by Eq.~(\ref{fitting-function-normalized}) focuses on the short-range part of the pair-correlation function
$g_{\uparrow \downarrow}(\rho)$ given by Eq.~(\ref{definition-pair-correlation-function}), which is dominated by the intra-pair correlations of relevance here.
It thus disregards a possible long-range part of $g_{\uparrow \downarrow}(\rho)$ which may include correlations between spin-$\uparrow$ and spin-$\downarrow$ fermions
belonging to different pairs (although this long-range part does not occur within the $t$-matrix approach adopted here).

\vspace{0.05cm}
\begin{center}
{\bf B. Pair fraction}
\end{center}

We are now in a position to calculate the pair density $n_{\mathrm{p}}$ given by
\begin{equation}
n_{\mathrm{p}} = - \int \! \frac{d\mathbf{q}}{(2 \pi)^{3}} \frac{1}{\beta} \sum_{\nu} e^{i \Omega_{\nu} \eta} \, \mathcal{G}_{B}(\mathbf{q},\Omega_{\nu}) 
\label{bosonic-density-homogeneous}
\end{equation}
together with the fermionic density $n_{\sigma}$ given by Eq.~(\ref{fermionic-density-equation}), for a homogeneous system as a function of coupling and temperature.

\begin{figure}[t]
\begin{center}
\includegraphics[width=8.0cm,angle=0]{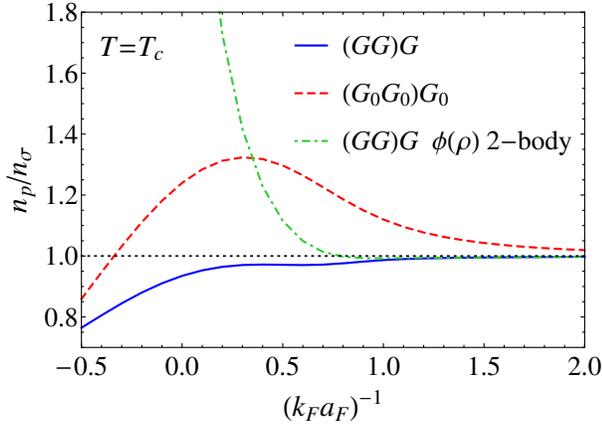}
\caption{(Color online) Pair fraction $n_{\mathrm{p}}/n_{\sigma}$ at $T_{c}$ vs $(k_{F} a_{F})^{-1}$, obtained by the fully self- consistent (full line) and non-self-consistent (dashed line) $t$-matrix approaches.
                                     In both cases, $\phi(\mathbf{p})$ in the form factors (\ref{def-F-j}) is obtained from the expression (\ref{fitting-function-normalized-FT}) within the respective approximations for the pair
                                     correlation function.
                                    Also shown is the result obtained by the fully self-consistent calculation, with $\phi(\mathbf{p})$ approximated instead by the two-body form (\ref{psi-k-BEC}) (dashed-dotted line).}
\label{Figure-3}
\end{center}
\end{figure} 

To begin with, Fig.~\ref{Figure-3} compares the pair fraction $n_{\mathrm{p}}/n_{\sigma}$ at $T_{c}$ over a wide range of the coupling $(k_{F} a_{F})^{-1}$, as obtained by
the fully self-consistent and non-self-consistent $t$-matrix approaches. 
As for other thermodynamic quantities \cite{PPS-2019}, also in this case the fully self-consistent approach proves superior to the non-self-consistent one, to the extent that the ratio $n_{\mathrm{p}}/n_{\sigma}$
should never exceed unity.
Accordingly, from now on results obtained by the fully self-consistent approach will only be presented. 
In addition, the use of the two-body form (\ref{psi-k-BEC}) for $\phi(\mathbf{p})$ in the form factors (\ref{def-F-j}) is seen to lead to unstable results upon entering the unitary regime with $(k_{F} a_{F})^{-1} \lesssim +1$.
Abandoning the two-body form (\ref{psi-k-BEC}) in favor of the expression (\ref{fitting-function-normalized-FT}) associated with the pair correlation function is thus expected to yield a definite improvement over the
theoretical analysis made in Ref.~\cite{Ulm-Cam-2019} when accounting for the experimental values of the pair fraction for the trapped system (cf.~Section~\ref{sec:comparison_with_experiment}-B below).

\begin{figure}[t]
\begin{center}
\includegraphics[width=9.0cm,angle=0]{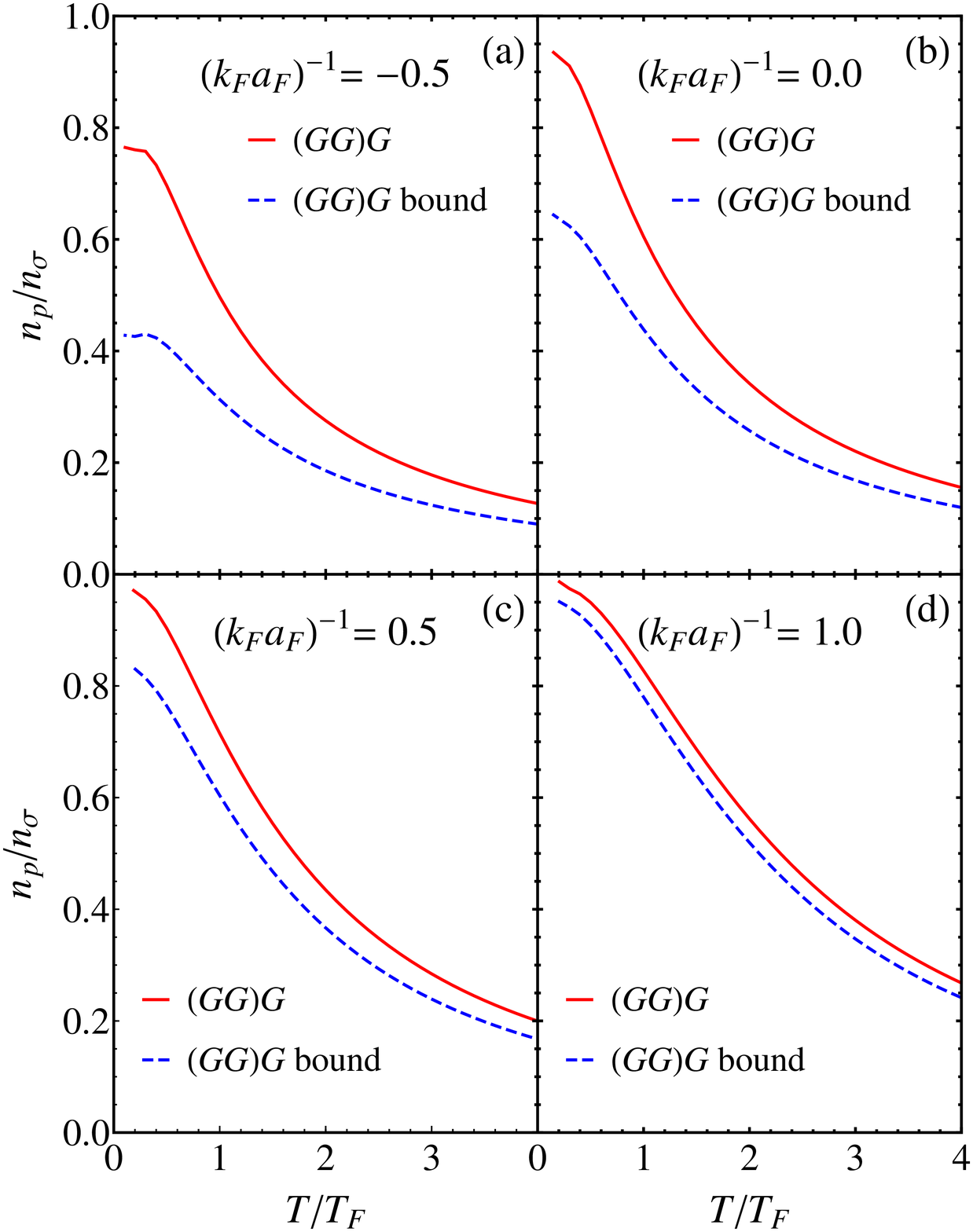}
\caption{(Color online) Pair fraction $n_{\mathrm{p}}/n_{\sigma}$ vs $T/T_{F}$ for four couplings, obtained by the fully self-consistent $t$-matrix approach including (full lines) 
                                    or neglecting (dashed lines) the ``unbound'' term in Eq.~(\ref{G_B-FT-sc}). In the latter case, only the ``bound'' term is retained in Eq.~(\ref{G_B-FT-sc}), as specified in the panels.}
\label{Figure-4} 
\end{center}
\end{figure} 

In Fig.~\ref{Figure-4} the pair fraction $n_{\mathrm{p}}/n_{\sigma}$ is shown over a wide range of temperature and a selected number of couplings across unitarity.
In particular, this figure compares the results obtained by including (full lines) or neglecting (dashed lines) the ``unbound'' term represented by the term - $\mathcal{F}_{2}$ on the right-hand side of Eq.~(\ref{G_B-FT-sc}).
One sees that inclusion of this unbound term over and above the bound term (represented by the second term on the right-hand side of Eq.~(\ref{G_B-FT-sc})) leads to substantial differences, especially in the unitary regime at low temperature.
The unbound term was not included in the diagrammatic  approach to the pair fraction presented in Ref.~\cite{Ulm-Cam-2019}.
It will be shown in Section~\ref{sec:comparison_with_experiment}-B that the agreement with experimental data will be definitively improved by its inclusion.

\begin{figure}[t]
\begin{center}
\includegraphics[width=9.0cm,angle=0]{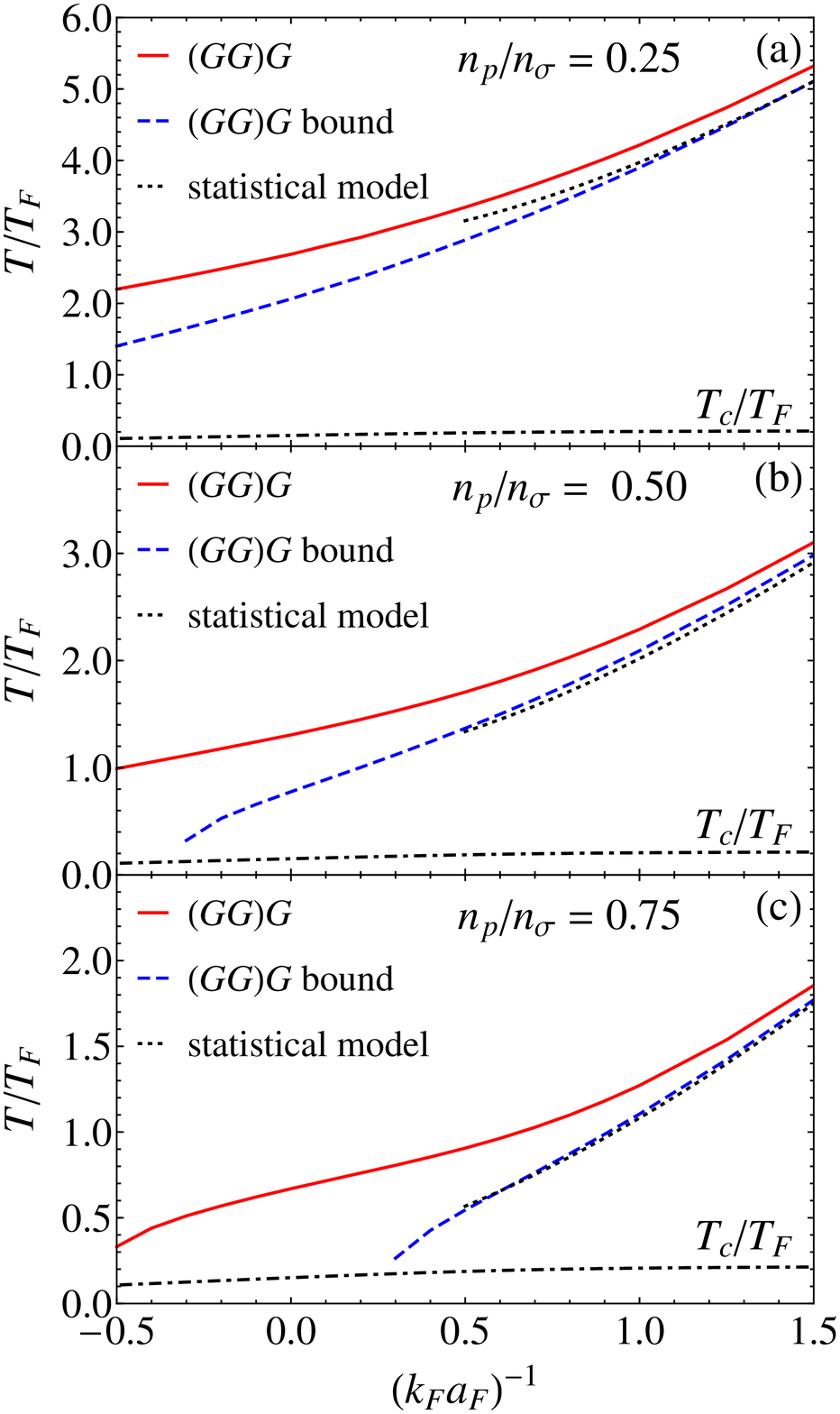}
\caption{(Color online) Contour plots of the pair fraction $n_{\mathrm{p}}/n_{\sigma}$ in the temperature-coupling phase diagram of the homogeneous system, obtained by the fully self-consistent 
                                    $t$-matrix approach by including (full lines) or neglecting (dashed lines) the ``unbound'' term in Eq.~(\ref{G_B-FT-sc}).
                                    Also shown are the results of the statistical model obtained from Eq.~(\ref{n_f-square-over-n_p-homogeneous}) (dotted lines).
                                    In each panel, the coupling dependence of the critical temperature $T_{c}$ (in units of $T_{F}$) is reported (dashed-dotted line), which sets the boundary of the normal phase 
                                    for the homogeneous system.}
\label{Figure-5}
\end{center} 
\end{figure} 

In preparation for this comparison, Fig.~\ref{Figure-5} shows three contour plots where a given value of the pair fraction $n_{\mathrm{p}}/n_{\sigma}$ is seen to evolve in 
the $T$-vs-$(k_{F} a_{F})^{-1}$ phase diagram.
Similarly to what was done in Fig.~\ref{Figure-4}, for each of the three values of $n_{\mathrm{p}}/n_{\sigma}$ here reported the numerical results have been obtained by including (full lines) or neglecting (dashed lines) the unbound term in Eq.~(\ref{G_B-FT-sc}).
In all cases, the difference between these two sets of results turns out to be substantial as soon as entering the unitary regime with $(k_{F} a_{F})^{-1} \lesssim +1$. 
This implies that, in this regime of most physical interest, the fermionic character of the constituent particles reveals itself.
As a consequence, this counting has to rely on filtering the occurrence of fermionic correlations, and not merely on signaling the presence of bound pairs which would instead apply to the molecular regime with $(k_{F} a_{F})^{-1} \gtrsim+1$.
 
To confirm this point of view, Fig.~\ref{Figure-5} also shows for comparison the contour plots of $n_{\mathrm{p}}/n_{\sigma}$ corresponding to the statistical model (dotted lines), as obtained from the law of mass action 
\begin{equation}
\frac{n_{\mathrm{f}}^{2}}{n_{\mathrm{p}}} = \frac{1}{8} \left( \frac{m k_{B} T}{\pi} \right)^{3/2} \!\! e^{-\varepsilon_{0}/k_{B}T} 
\label{n_f-square-over-n_p-homogeneous}
\end{equation}
where $\varepsilon_{0} = (m a_{F}^{2})^{-1}$ is the two-body binding energy, which results from the integrals in Eq.~(\ref{density-decomposition}) of Appendix \ref{sec:appendix-model} 
by neglecting $\pm 1$ in the denominators therein.
It turns out that the results of the statistical model coincides with those of the quantum many-body approach that includes only the bound term, but only at most up to $(k_{F} a_{F})^{-1} \approx 0.6$ after which the molecular regime with the two-body wave function (\ref{psi-rho-BEC}) loses its meaning.

\section{Results for a trapped gas and comparison with experimental data} 
\label{sec:comparison_with_experiment}

The results obtained in Section~\ref{sec:numerical_results_homogeneous} for $n_{\mathrm{p}}$ given by Eq.~(\ref{bosonic-density-homogeneous}) 
and for $n_{\sigma}$ given by Eq.~(\ref{fermionic-density-equation}) refer to a homogeneous system.
In order to compare with the experimental data of Ref.~\cite{Ulm-Cam-2019}, these theoretical results need to be averaged over the trap that contains the Fermi gas.

\vspace{0.05cm}
\begin{center}
{\bf A. Trap average}
\end{center}

\begin{figure}[t]
\begin{center}
\includegraphics[width=9.0cm,angle=0]{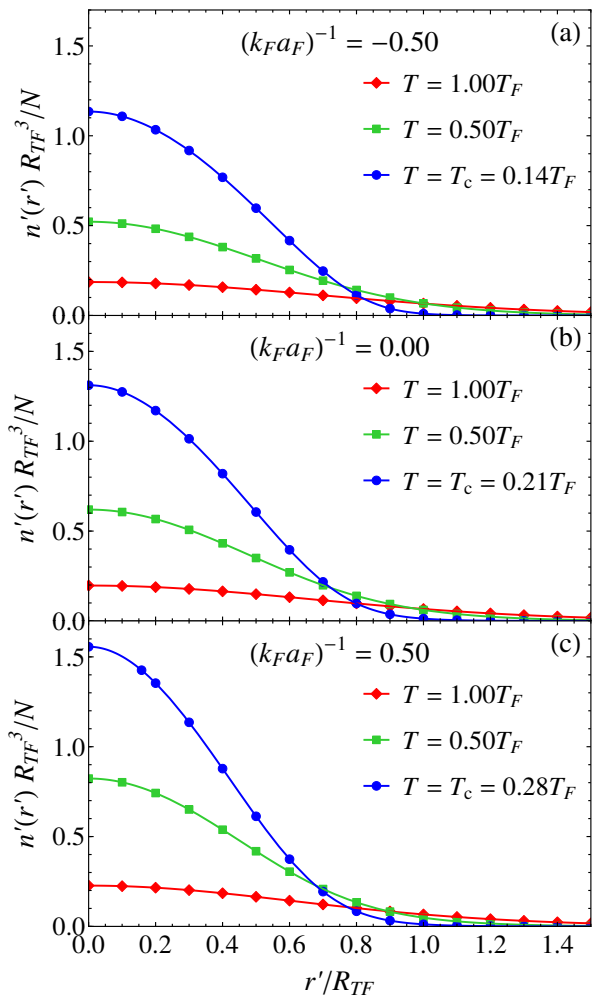}
\caption{(Color online) Isotropic radial density $n'(r')$ vs $r'$ for couplings: (a) $(k_{F} a_{F})^{-1} = -0.5$; (b) $(k_{F} a_{F})^{-1} = 0.0$; (c) $(k_{F} a_{F})^{-1} = +0.5$. 
                                    In each panel, the results for $T=T_{c}$ (dots), $T=0.5T_{F}$ (squares), $T=T_{F}$ (diamonds) are shown.
                                    Lengths are in units of the Thomas-Fermi radius $R_{TF}$ given by $\frac{1}{2} m \omega_{0}^{2} R_{TF}^{2} = E_{F}$
                                    where $E_{F} = \omega_{0} (3N)^{1/3}$ is the trap Fermi energy, such that  $8N/(\pi^2 R_{TF}^3)$ is the value of $n(r=0)$ for the non-interacting 
                                    gas at $T=0$ within a local-density approximation.}
\label{Figure-6}
\end{center}
\end{figure} 

When considering a Fermi gas trapped in an anisotropic harmonic potential of the type
\begin{equation}
V(\mathbf{r}) = \frac{1}{2} m \left( \omega_{x}^{2} x^{2} + \omega_{y}^{2} y^{2} + \omega_{z}^{2} z^{2} \right) \, ,
\label{anisotropic-harmonic-potential-fermions}
\end{equation}
one can adopt a \emph{local-density approach} and obtain the total number $N_{\mathrm{p}}$ of pairs and the total number $N_{\sigma}$ of fermions in the trap in the following way.
One first replaces the fermionic chemical potential $\mu$ entering the single-particle Green's function $\mathcal{G}(\mathbf{p},\omega_{n})$ of Eq.~(\ref{self-consistent-G})
by $\mu \rightarrow \mu - V(\mathbf{r})$, thereby obtaining the \emph{local} function $\mathcal{G}(\mathbf{p},\omega_{n};\mathbf{r})$.
One then replaces $\mathcal{G}(\mathbf{p},\omega_{n}) \rightarrow \mathcal{G}(\mathbf{p},\omega_{n};\mathbf{r})$ everywhere this function occurs, namely, in the expressions (\ref{G_B-FT-sc})-(\ref{pp-bubble}) for pairs and the expressions (\ref{self-consistent-G})-(\ref{fermionic-density-equation}) for fermions.
Finally, one integrates the expressions of the local densities $n_{\mathrm{p}}(\mathbf{r})$ and $n_{\sigma}(\mathbf{r})$ obtained in this way over the spatial variable $\mathbf{r}$, to get the total number of pairs $N_{\mathrm{p}}$ and the total number of fermions $N_{\sigma}$ with spin $\sigma$.
The value of the fermionic chemical potential $\mu$ \emph{for the trap} is eventually determined for given coupling and temperature by solving for $\mu$ as a function of $N_{\sigma}$.
In practice, in the experiment of Ref.~\cite{Ulm-Cam-2019} typical values of $\omega_{x} = \omega_{y}$ range from $2 \pi \times 300$ Hz to $2 \pi \times 1.6$ kHz, while
$\omega_{z} = \lambda \omega_{x} = 2 \pi \times 21$ Hz (with $\lambda < 1$).

In the theoretical expressions, it is convenient to map at the outset the anisotropic potential (\ref{anisotropic-harmonic-potential-fermions}) into a spherical one by rescaling the variables
from $(x,y,z)$ to $(x' = \lambda^{-1/3} x$,$y' = \lambda^{-1/3} y$,$z' = \lambda^{2/3} z)$, such that the trapping potential becomes
\begin{equation}
V(x',y',z') = \frac{1}{2} m \, \omega_{0}^{2} \, r'^{2}
\label{isotropic-harmonic-potential-fermions}
\end{equation} 
where $r' = \sqrt{x'^{2} + y'^{2} + z'^{2}}$ and $\omega_{0} = (\omega_{x} \omega_{y} \omega_{z})^{1/3} = \lambda^{1/3} \omega_{x}$ is the average trap frequency.
Accordingly, the original spatial distribution $n(x,y,z)$ of the fermionic density with an ellipsoidal shape is mapped onto a spherical distribution $n'(x',y',z') = n'(r')$ through the rescaling 
$n(x,y,z) = n'(\lambda^{-1/3}x,\lambda^{-1/3}y,\lambda^{2/3}z)$ (where both spin components are meant to be included).

Profiles of the total fermionic isotropic density $n'(r')$ obtained in this way are shown in Fig.~\ref{Figure-6}, for several couplings across unitarity and temperatures in the normal phase.
The coupling parameter $(k_{F} a_{F})^{-1}$ associated with the trap is expressed in terms of $k_{F} = \sqrt{2 m E_{F}}$, where $E_{F} = \omega_{0} (3N)^{1/3}$ is the Fermi energy of the trap and $N=N_{\uparrow}+N_{\downarrow}$ is the total number of fermions.
(In the experiment of Ref.~\cite{Ulm-Cam-2019}, typical values of $N$ range from $3 \times 10^4$ to $3 \times 10^5$.)

The values of the critical temperature $T_{c}$ for the trap case, reported in Fig.~\ref{Figure-6} only for three specific couplings, can be obtained throughout the whole BCS-BEC crossover.
This information is important also to verify whether the experimental values of the pair fraction in the trap of Ref.~\cite{Ulm-Cam-2019} were measured in the normal phase.
To calculate $T_{c}$ for the trap, we adopt again a local-density approach and define a local Fermi temperature $T_{F}(\mathbf{r})$ such that 
$k_{B} T_{F}(\mathbf{r}) = \left[ 3 \pi^{2} n(\mathbf{r})\right]^{2/3}\!/(2m)$.
This implies that the local Fermi temperature, like the density $n(\mathbf{r})$, has its maximum value at $\mathbf{r}=0$, to which there corresponds a minimum value of $T/T_{F}(\mathbf{r})$ for given temperature $T$.
Accordingly, the central portion of the cloud density is where superfluidity is first established upon lowering the temperature from the normal phase.

To obtain $T_{c}$ for the trapped system, we then apply the Thouless criterion
\begin{equation}
\Gamma \left( \mathbf{q}=0,\Omega_{\nu}=0;\mu(\mathbf{r}=0),T_{c} \right)^{-1} = 0
\label{local-Thouless-criterion}
\end{equation}
in terms of the particle-particle propagator (\ref{Gamma}) in the normal phase, where now $\mu(\mathbf{r}=0) = \mu - V(\mathbf{r}=0) = \mu$ is the fermionic chemical potential for the trap calculated
at the critical temperature $T_{c}$.
Details on how the variables $(T_{c},\mu_{c})$ have been determined by solving the Thouless criterion in conjunction with the density equation are given in Appendix B of Ref.~\cite{PPS-2019}.

\begin{figure}[t]
\begin{center}
\includegraphics[width=8.0cm,angle=0]{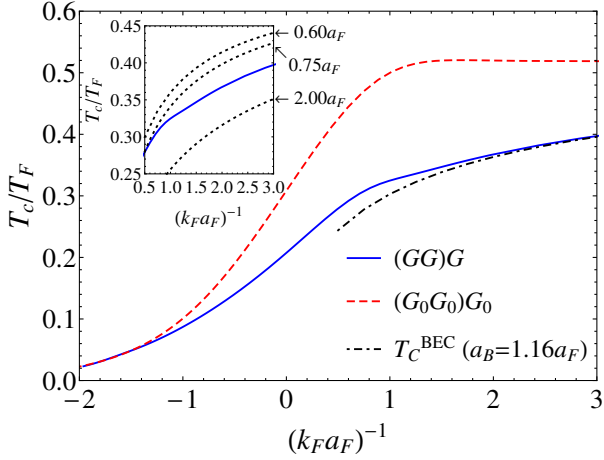}
\caption{(Color online) Critical temperature $T_{c}$ (in units of the Fermi temperature $T_{F} = E_{F}/k_{B}$) vs $(k_{F} a_{F})^{-1}$ for the trapped system.
                                    Results are shown for the fully self-consistent (full line) and for the non-self-consistent (dashed line) $t$-matrix approaches.
                                    In the BEC regime, the results of a model calculation for trapped bosons with a mean-field-type interaction (cf.~Appendix C) are also shown with the value $a_{B} = 1.16 a_{F}$
                                    for the bosonic scattering length (dashed-dotted line).
                                    The inset shows the results of additional bosonic calculations with different values of $a_{B}$ (see text).}
\label{Figure-7}
\end{center}
\end{figure} 

Figure~\ref{Figure-7} shows the results of our calculation for $T_{c}$ in the trap across the BCS-BEC crossover.
The results of the fully self-consistent $t$-matrix approach (full line) are also compared with those of its non-self-consistent counterpart (dashed line).
While the two calculations essentially coincide with each other in the BCS regime $(k_{F} a_{F})^{-1} \lesssim -1$, they differ considerably on the BEC side of unitarity.
We attribute this difference to the occurrence of a residual interaction between composite bosons in the BEC regime $(k_{F} a_{F})^{-1} \gtrsim +1$, which is present within the fully self-consistent 
but absent within the non-self-consistent calculation \cite{PPS-2019}.

To make a check on the results of our numerical calculation, also shown in Fig.~\ref{Figure-7} are the results for $T_{c}$ (dashed-dotted line) 
obtained for a low-density trapped Bose gas with a residual interaction specified by the scattering length $a_{B}$ (cf.~Appendix \ref{sec:appendix-scattering-length}), where for internal consistency
the (approximate) value $a_{B} = 1.16 a_{F}$ that results from the fully self-consistent $t$-matrix approach \cite{PPS-2019} was considered.
In this way, we can confirm quantitatively the effects of $a_{B}$ on $T_{c}$ for the trapped system in the BEC regime, which are contained in the fully self-consistent $t$-matrix approach.
For comparison, the inset reports additional bosonic calculations for: (i)~$a_{B} = 2.0 a_{F}$ which corresponds to the residual bosonic interaction being treated at the level of the fermionic exchange diagrams \cite{PS-2000}; (ii)~$a_{B} = 0.75 a_{F}$ when the $T$-matrix for the dimer-dimer scattering built on these exchange diagrams is further considered \cite{PS-2000}; (iii) The exact value $a_{B} = 0.6 a_{F}$ obtained either by a numerical solution of the four-body Schr\"{o}dinger equation \cite{Petrov-2005} or by a full diagrammatic treatment in the zero-density limit \cite{Brodsky-2006}.

Finally, it should be mentioned that the value $T_{c}/T_{F} = 0.2074$, which we have obtained at unitarity by the fully self-consistent calculation, coincides with that obtained in Ref.~\cite{Haussmann-2008} by the same approach.
However, our calculation for $T_{c}$ is extended to the whole BCS-BEC crossover while that of Ref.~\cite{Haussmann-2008} was limited to unitarity only.

\vspace{0.05cm}
\begin{center}
{\bf B. Comparison between theory and experiment}
\end{center}

\begin{figure}[t]
\begin{center}
\includegraphics[width=8.5cm,angle=0]{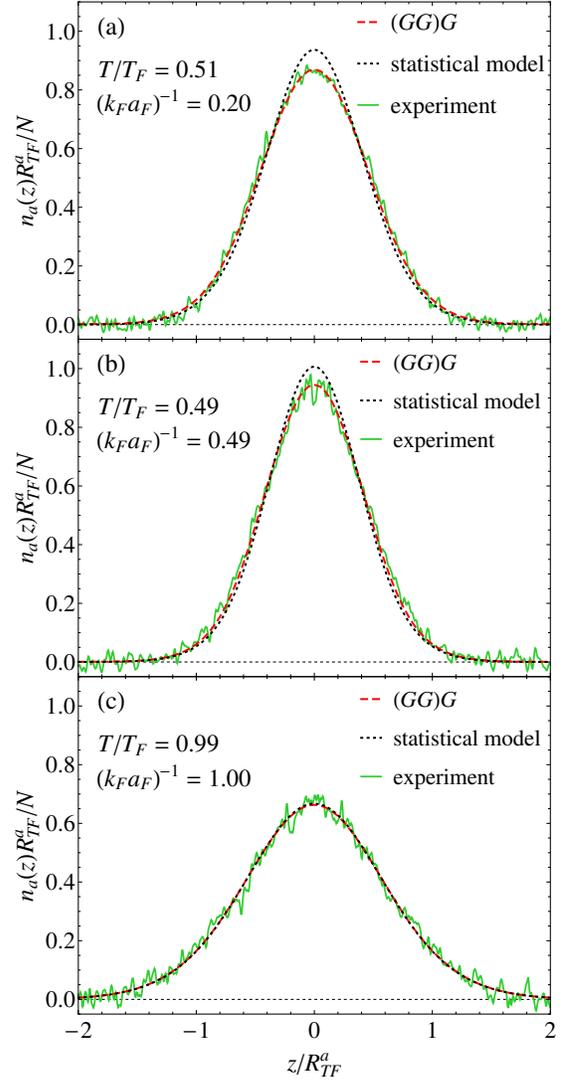}
\caption{(Color online) Comparison between the axial densities along the main axis of the trap, as observed experimentally (full lines) and calculated with the self-consistent $t$-matrix approach 
                                    (dashed lines) and the statistical atom-molecule model (dotted lines), when (a) $T/T_{F} = 0.51(4)$ and $(k_{F} a_{F})^{-1} = 0.20(3)$, (b) $T/T_{F} = 0.49(4)$ and 
                                    $(k_{F} a_{F})^{-1} = 0.49(3)$, and (c) $T/T_{F} = 0.99(6)$ and $(k_{F} a_{F})^{-1} = 1.00(5)$. 
                                     The ratio $\lambda$ between the axial and radial trap frequencies equals $0.0435(8)$ in (a), $0.0424(7)$ in (b), and $0.0272(4)$ in (c).
                                     The axial Thomas-Fermi radius $R^a_{TF}= \lambda^{-2/3} \, R_{TF}$ is used for normalization.}
\label{Figure-8}
\end{center}
\end{figure} 

A first quantity to be compared with the experimental data of Ref.~\cite{Ulm-Cam-2019} is the so-called \emph{axial} density $n_{\mathrm{a}}(z)$ where $z$ runs along the main axis of the trap, 
which is obtained by integrating the full density $n(x,y,z)$ over the radial directions $x$ and $y$.
Specifically, the experimental profiles $n_{\mathrm{a}}(z)$ can be compared with their theoretical counterparts $n'_{\mathrm{a}}(z')$, obtained by integrating over $x'$ and $y'$ the isotropic profiles $n'(r') = n'(x',y',z')$ (like those shown in Fig.~\ref{Figure-6}) and then performing the rescaling
\begin{equation}
n_{\mathrm{a}}(z) \, = \, \lambda^{2/3} \, n'_{\mathrm{a}}(\lambda^{2/3} z) \, .
\label{rescaling}
\end{equation} 
Figure~\ref{Figure-8} shows this comparison for three sets of values of temperature, coupling, and anisotropy $\lambda$.
In all cases, excellent agreement results between the experiment and the quantum many-body approach with no adjustable parameter.
The figure shows also the comparison with the statistical atom-molecule model, for which notable deviations from the experiment occur, as expected, for low temperature and close to unitarity.

\begin{figure}[t]
\begin{center}
\includegraphics[width=9.0cm,angle=0]{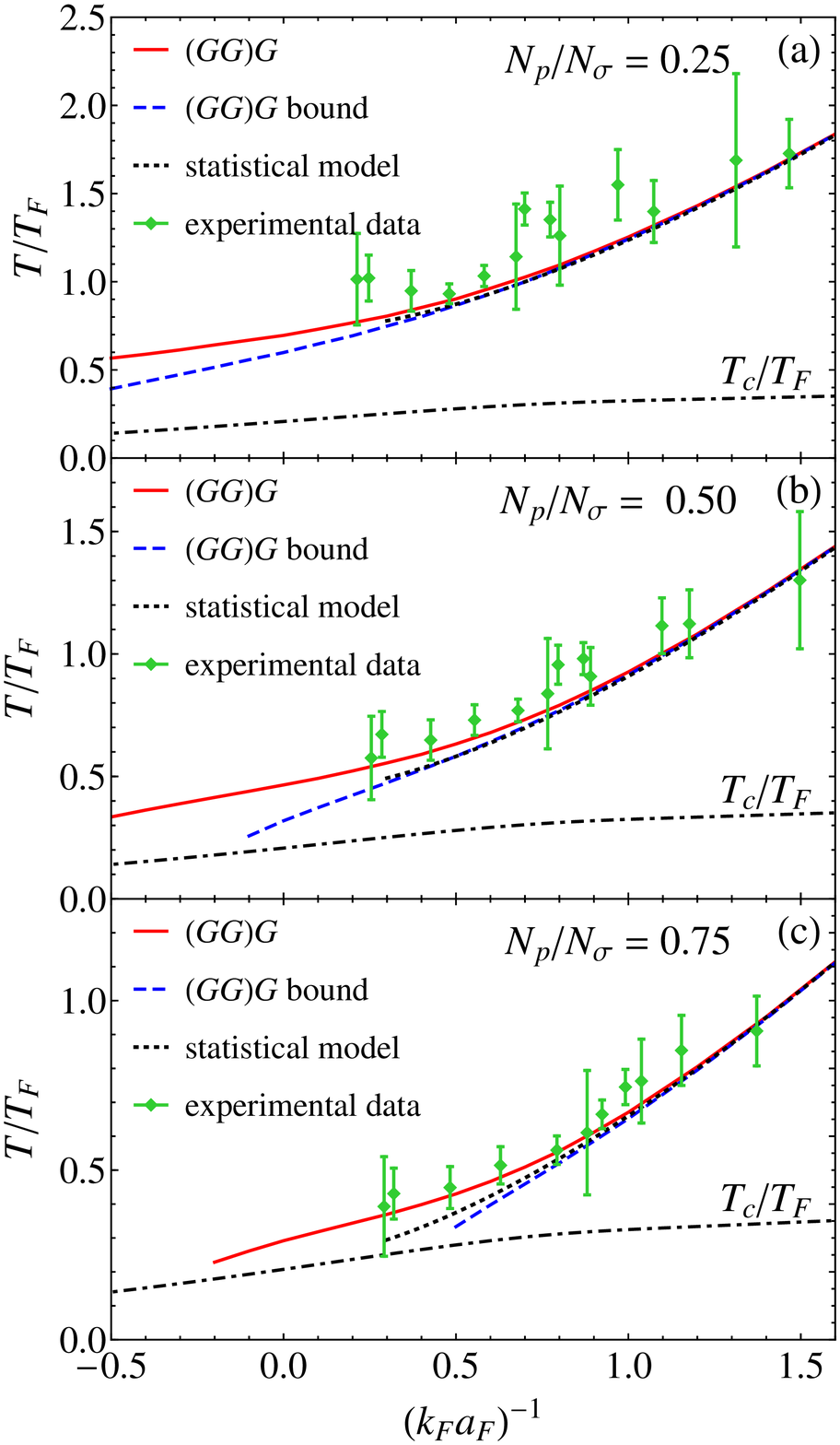}
\caption{(Color online) Contour plots of the pair fraction $N_{\mathrm{p}}/N_{\sigma}$ in the temperature-coupling phase diagram of the trapped system, obtained by the fully self-consistent 
                                    $t$-matrix approach by including (full lines) or neglecting (dashed lines) the ``unbound'' term in Eq.~(\ref{G_B-FT-sc}).
                                    The theoretical curves are compared with the experimental data of Ref.~\cite{Ulm-Cam-2019} (diamonds with vertical error bars).
                                    For $(k_{F} a_{F})^{-1} \gtrsim 0.3$ the results of the statistical model obtained from Eq.~(\ref{N_f-square-over-N_p-trapped}) of 
                                    Appendix~\ref{sec:appendix-model} are also shown (dotted lines).
                                    In each panel, the coupling dependence of the critical temperature $T_{c}$ (in units of $T_{F}$) is reported (dashed-dotted line), which sets the boundary of the normal phase 
                                    for the trapped system.}
\label{Figure-9}
\end{center}
\end{figure} 

Finally, Fig.~\ref{Figure-9} presents the comparison of the pairing fraction $N_{\mathrm{p}}/N_{\sigma}$ obtained by our \emph{ab initio} quantum many-body calculation with the experimental data of  Ref.~\cite{Ulm-Cam-2019} over the temperature-coupling phase diagram (where $k_{F}$ and $T_{F}$ now refer to the trapped system). The comparison is made for three characteristic values of $N_{\mathrm{p}}/N_{\sigma}$.
In all cases, good agreement is obtained between theory and experiment (we emphasize that the theoretical results have been obtained with no adjustable parameter).

In particular, this comparison shows that the contribution of the unbound term significantly improves the agreement of our calculations with the experimental data, despite the presence of the trap which acts to suppress the contribution of the unbound term (which is evident by comparing Figs.~\ref{Figure-5} and \ref{Figure-9}). This suggests that the experimental data probe indeed the pairing \emph{correlations} between spin-up and spin-down fermions as defined by our formalism. 

From this comparison one can argue that the crossover, between the pseudo-gap regime (where the fermionic character of the constituent particles matters) and the molecular regime (where only the presence of bosonic pairs is relevant), sets in about where the theoretical results for $N_{\mathrm{p}}/N_{\sigma}$, obtained with and without the unbound term, start departing from each other.
This argument cannot be made in terms of the statistical atom-molecule model \cite{Ulm-Cam-2019}, that misses the contribution of the unbound term.

\section{Concluding remarks}
\label{sec:conclusions}

In this paper, we have provided a detailed account of a theoretical approach to interpret the experimental data reported in Ref.~\cite{Ulm-Cam-2019} in a quantitative way.
By this approach, from the data Ref.~\cite{Ulm-Cam-2019} we have been able to unravel how the occurrence of pairing correlations between spin-up and spin-down fermions at equilibrium develops, as a function of temperature in the normal phase and of coupling on the BEC side of unitarity.
What we claim to have learned from this is how the pseudo-gap regime (where fermions matter) and the molecular regime (where only composite bosons matter) separate from each other. This should be considered rather remarkable, since this result was extracted from experiment \cite{Ulm-Cam-2019} where an equilibrium quantity was measured (i.e.~the number of fermion pairs) and not a dynamical quantity (the excitation gap).
 
From the theoretical side, to account for the experimental data we have taken advantage of several favorable circumstances.
On the one hand, since the number of fermion pairs in a Fermi gas undergoing the BCS-BEC crossover is an equilibrium quantity, it can be accounted for quite well in terms of the fully self-consistent $t$-matrix approach \cite{PPS-2019}.
On the other hand, this physical quantity that was measured experimentally by its own nature does not require one to endow the theory with a series of complicated Aslamazov-Larkin and Maki-Thompson diagrams, 
which should otherwise be included to fulfill conservation criteria when addressing dynamical response functions \cite{Baym-1962}, to the extent that the single-particle self-energy is treated within the fully self-consistent $t$-matrix approach.
In addition, our emphasis here on fermionic correlations has drawn on our previous experience on the pair-correlation function in the normal phase, which was addressed in detail in Ref.~\cite{Palestini-2014}
within the non-self-consistent $t$-matrix approach and here extended to the fully self-consistent one.

Along these lines, future perspectives, that could reinforce our argument about the evidence for the separation between the (fermionic) pseudo-gap and the (bosonic) molecular regimes, may hinge on the possibility of extending the measurements of the ratio $N_{\mathrm{p}}/N_{\sigma}$ towards unitarity at temperatures close enough to $T_{c}$.

In addition, to highlight experimentally the relevance of the correlations induced indirectly by the environment between spin-$\uparrow$ and spin-$\downarrow$ fermions, which are embodied in the ``unbound'' term in the expression (\ref{G_B-FT-sc}), it could be worth to consider repeating the experiment of  Ref.~\cite{Ulm-Cam-2019} by replacing the harmonic trap with a box trap along the lines of Ref.~\cite{Zwierlein-2017}. 
In this way, one should be able to amplify the difference between the values of the pair fraction obtained with and without the inclusion of the unbound term, as one may anticipate by 
comparing the results of Fig.~\ref{Figure-5} for the homogeneous case with those of Fig.~\ref{Figure-9} for the trapped case.

It is, finally, interesting to draw a physical connection between our finding about the indirect correlations established between spin-$\uparrow$ and spin-$\downarrow$ fermions through their environment and the recent results of Ref.~\cite{CLAS-2019} about the way the quark-gluon structure of a nucleon bound in an atomic nucleus is modified by the surrounding nucleons.
In both cases, it is the environment that plays an important role in modifying the properties of what would be a bound system in isolation.


\begin{center}
\begin{small}
{\bf ACKNOWLEDGMENTS}
\end{small}
\end{center}

MP, PP, and GCS acknowledge financial support from the Italian MIUR under Projects PRIN2015 (2015C5SEJJ001) and PRIN2017 (CEnTraL 20172H2SC4).
MJ and JHD acknowledge financial support from DFG (LI988/6-1), and thank W. Limmer, T. Paintner and D. Hoffmann for discussions.


\appendix   

\section{ABOUT THE USE OF CONSERVING APPROXIMATIONS FOR THE PAIR FRACTION}
\label{sec:appendix-conserving}

In Section \ref{sec:theoretical_approach}-B we have argued that \emph{only} the form (\ref{csibcs}) of the effective two-particle interaction $\Xi$ is of relevance for the calculation of the bosonic propagator
$\mathcal{G}_{B}(\mathbf{q},\Omega_{\nu})$ of Eq.~(\ref{G_B-FT}) (and thus of the quantity $N_{\mathrm{p}}$ of experimental interest).
We have also anticipated that the reason for this is to be found in the specific sequence of Nambu indices appearing in the expression (\ref{G-L}) from which Eq.~(\ref{G_B-FT}) is derived.
Here, we show specifically how the diagrammatic contributions to $\Xi$, that would derive from the $t$-matrix approach for the fermionic self-energy $\Sigma$, cannot modify this result.
Under different circumstances, like for the calculation of the density and spin response functions, on the other hand, the diagrams for $\Xi$ corresponding to the Aslamazov-Larkin (AL) and Maki-Thomson (MT) 
contributions would instead result from the $t$-matrix approach for $\Sigma$ (see, e.g., Fig. 3 of Ref.~\cite{SPL-2002}).
In our case, the importance of introducing the $t$-matrix approach for $\Sigma$ arises from the need of obtaining an accurate description of the thermodynamic properties of the Fermi gas in the normal phase \cite{PPS-2019}.

\begin{figure}[t]
\begin{center}
\includegraphics[width=8.5cm,angle=0]{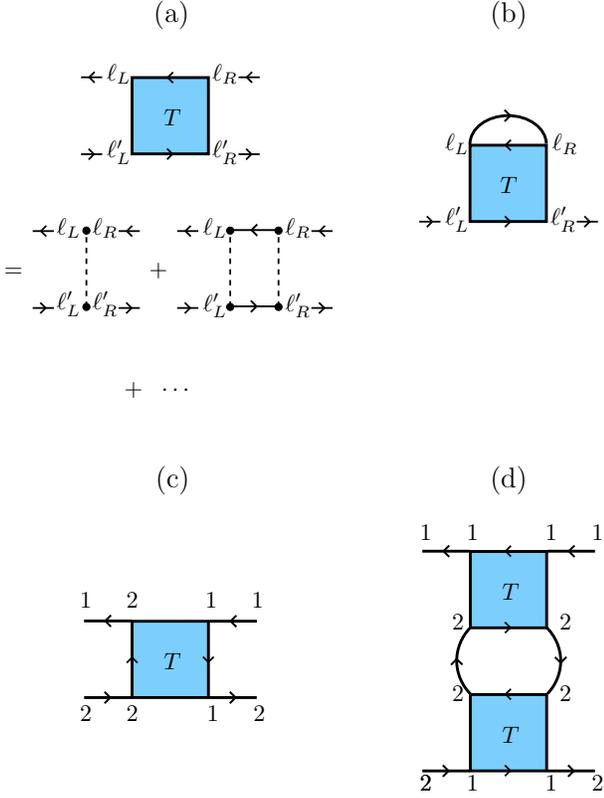}
\caption{(Color online) 
(a) Ladder diagrams for the T-matrix in the superfluid phase, where dots delimiting potential (dashed) lines represent $\tau^{3}$ Pauli matrices.
(b) Corresponding diagram for the $t$-matrix fermionic self-energy.  Examples of (c) MT and (d) AL diagrammatic contributions to the pair propagator $\mathcal{G}_{B}$, 
 which are bound to vanish when carried over to the normal phase owing to the presence of two anomalous fermionic single-particle Green's functions which connect $1 \leftrightarrow 2$.
For simplicity, only Nambu spin indices have been explicitly indicated in all diagrams.}
\label{Figure-10}
\end{center}
\end{figure} 

Probably the simplest way to convince oneself that the AL-type and MT-type contributions to $\Xi$, which would result from the $t$-matrix self-energy taken below $T_{c}$, do not contribute to the expression (\ref{G-L}) of the pair propagator $\mathcal{G}_{B}$ once carried over to the normal phase above $T_{c}$, is to draw these contributions in a diagrammatic way.
This is done in Fig.~\ref{Figure-10}.
Here, the series of ladder diagrams that approximate the many-particle T-matrix in the broken-symmetry phase is reported in panel (a), while the corresponding $t$-matrix self-energy is shown 
in panel (b). 
For simplicity,  in these diagrams only the Nambu indices have been explicitly indicated, while the space and imaginary time variables are not reported since they are not essential to the following argument.
The crucial point is that for the T-matrix of panel (b) only combinations with Nambu indices $\ell_{L} \ne \ell'_{L}$ and $\ell_{R} \ne \ell'_{R}$ occur, owing to the inter-particle interaction of the contact form that we have adopted
(cf.~also Ref.~\cite{Andrenacci-2003}).
In addition, only combinations with $\ell_{L} = \ell_{R}$ and $\ell'_{L} = \ell'_{R}$ will survive when these diagrams are extrapolated to the normal phase.
As a consequence, a typical example of MT contribution is shown in Fig.~\ref{Figure-10}(c), while a typical example of AL contribution is shown in Fig.~\ref{Figure-10}(d).
In all cases, it turns out that at least two single-particle Green's functions with off-diagonal Nambu indices would be required to match these contributions to $\Xi$ with the Nambu indices appearing in the expression 
(\ref{G-L}).
Since the off-diagonal (anomalous) single-particle Green's functions vanish in the normal phase above $T_{c}$, the MT- and AL-type contributions to $\Xi$ vanish, too, and do not affect
the expression (\ref{G-L}) which is relevant for the calculation of $N_{\mathrm{p}}$ above $T_{c}$.
This proves our statement.

\section{COMPARISON BETWEEN THE QUANTUM MANY-BODY APPROACH AND THE STATISTICAL ATOM-MOLECULE MODEL FOR THE PAIR FRACTION}
\label{sec:appendix-model}

It is interesting to determine under what physical circumstances the expressions for the total number of bosons $N_{\mathrm{p}}$ and for the total number of spin-$\sigma$ fermions $N_{\sigma}$ of our fully quantum many-body approach reduce to those of a statistical model of a fermion-boson mixture at equilibrium 
\cite{Chin-Grimm-2004,Eagles-1969}.

To this end, we consider a homogeneous system, for which $N_{\mathrm{p}} = \mathcal{V} \, n_{\mathrm{p}}$ and $N_{\sigma} = \mathcal{V} \, n_{\sigma}$ are expressed in terms of the respective densities.
By our quantum many-body approach, $n_{\mathrm{p}}$ is given by Eq.~(\ref{bosonic-density-homogeneous}) with $\mathcal{G}_{B}$ given by the expression (\ref{G_B-FT-sc}), 
while $n_{\sigma}$ is given by the expression (\ref{fermionic-density-equation}).
To recover the physics of a fermion-boson (or atom-molecule) mixture, one requires the fermionic coupling to be sufficiently strong in the BEC regime and the temperature sufficiently low, for the internal structure of the composite bosons (dimers) to become irrelevant.

In this limit, the fermionic chemical potential $\mu$ becomes the largest energy scale of the problem and is written in the form $\mu_{B} = 2 \mu + \varepsilon_{0}$, where $\varepsilon_{0}= (m a_{F}^{2})^{-1}$ is the dimer binding energy and $\mu_{B}$ the dimer chemical potential \cite{Physics-Reports-2018}.
The expression (\ref{Pi}) then reduces to
\begin{equation}
\tilde{\Pi}(\mathbf{p};\mathbf{q},\Omega_{\nu}) \simeq \frac{1}{ 2 \xi(\mathbf{p})} \, ,
\label{Pi-BEC_limit}
\end{equation}
which, together with the expression (\ref{psi-k-BEC}) for $\phi({\mathbf p})$ appropriate to this limit, yields the the following approximate form for the form factors (\ref{def-F-j}) \cite{Andrenacci-2003}:
\begin{equation}
\mathcal{F}_{1}(\mathbf{q},\Omega_{\nu}) \simeq  \sqrt{\frac{m^{2} a_{F}}{8 \pi}} \hspace{0.3cm} , \hspace{0.3cm} \mathcal{F}_{2}(\mathbf{q},\Omega_{\nu}) \simeq  \frac{m a_{F}^{2}}{4} \, .
\label{approximate-form-factors}
\end{equation}
This implies that, in the BEC limit where $a_F \to 0^+$, the ``unbound'' term $\mathcal{F}_{2}$ vanishes faster than $\mathcal{F}_{1}$ and can thus be neglected in the expression (\ref{G_B-FT-sc}).
In addition, in the same limit the particle-particle propagator $\Gamma(\mathbf{q},\Omega_{\nu})$ of the ``bound'' term in the expression (\ref{G_B-FT-sc}) acquires the polar form \cite{Physics-Reports-2018}:
\begin{equation}
\Gamma(\mathbf{q},\Omega_{\nu}) \simeq - \frac{8 \pi}{m^{2} a_{F}} \, \frac{1}{i\Omega_{\nu} - \frac{\mathbf{q}^{2}}{4m} + \mu_{B}} \, .
\label{polar-particle-particel-propagator}
\end{equation}
Combining these results together, one gets eventually for the bosonic density:
\begin{eqnarray}
n_{\mathrm{p}} & \simeq & - \int \! \frac{d\mathbf{q}}{(2 \pi)^{3}} \frac{1}{\beta} \sum_{\nu} \frac{e^{i \Omega_{\nu} \eta}}{i\Omega_{\nu} - \frac{\mathbf{q}^{2}}{4m} + \mu_{B}} 
\nonumber \\
& = & \int \! \frac{d\mathbf{q}}{(2 \pi)^{3}} \, \frac{1}{e^{\beta \xi_{B}(\mathbf{q})} - 1}
\label{approximate-bosonic-density}
\end{eqnarray}
in terms of the Bose-Einstein distribution of argument $\xi_{B}(\mathbf{q}) = \frac{\mathbf{q}^{2}}{4m} - \mu_{B}$.

To determine $n_{\sigma}$ in the BEC limit at sufficiently low temperature, we consider the expression (\ref{fermionic-density-equation}) where we expand the single-particle Green's function (\ref{self-consistent-G}) in series of the self-energy $\Sigma$
\begin{equation}
\mathcal{G}(\mathbf{p},\omega_{n}) \simeq \mathcal{G}_{0}(\mathbf{p},\omega_{n}) + \mathcal{G}_{0}(\mathbf{p},\omega_{n}) \, \Sigma(\mathbf{p},\omega_{n}) \, \mathcal{G}_{0}(\mathbf{p},\omega_{n}) \, + \cdots
\label{G-expanded}
\end{equation}
where $\mathcal{G}_{0}(\mathbf{p},\omega_{n}) = [i \omega_{n} - \xi(\mathbf{p})]^{-1}$ is the non-interacting single-particle Green's function, 
by again relying on the fact that the fermionic chemical potential $\mu$ entering $\xi(\mathbf{p}) = \mathbf{p}^{2}/(2m) - \mu$ is the largest energy scale in the problem.
We thus obtain:
\begin{eqnarray}
n_{\sigma} & \simeq & \int \! \frac{d\mathbf{p}}{(2 \pi)^{3}} \frac{1}{\beta} \sum_{n} e^{i \omega_{n} \eta} \, \mathcal{G}_{0}(\mathbf{p},\omega_{n})  
\nonumber \\
& + & \int \! \frac{d\mathbf{p}}{(2 \pi)^{3}} \frac{1}{\beta} \sum_{n} \, \mathcal{G}_{0}(\mathbf{p},\omega_{n})^{2} \, \Sigma(\mathbf{p},\omega_{n}) \, + \, \cdots
\nonumber \\
& \equiv & n_{\sigma}^{(0)} \, + \, n_{\sigma}^{(1)} \, .
\label{approximate-fermionic-density-equation}
\end{eqnarray}
Here,
\begin{eqnarray}
n_{\sigma}^{(0)} & = & \int \! \frac{d\mathbf{p}}{(2 \pi)^{3}} \frac{1}{\beta} \sum_{n} e^{i \omega_{n} \eta} \, \mathcal{G}_{0}(\mathbf{p},\omega_{n}) 
\nonumber \\
& = &  \int \! \frac{d\mathbf{p}}{(2 \pi)^{3}} \, \frac{1}{e^{\beta \xi(\mathbf{p})} + 1}
\label{fermionic-density-zero}
\end{eqnarray}
coincides with the density $n_{\mathrm{f}}$ of fermions (atoms) expressed in terms of the Fermi-Dirac distribution of argument $\xi(\mathbf{p})$, and
\begin{eqnarray}
n_{\sigma}^{(1)} & = & \int \! \frac{d\mathbf{p}}{(2 \pi)^{3}} \frac{1}{\beta} \sum_{n} \mathcal{G}_{0}(\mathbf{p},\omega_{n})^{2} \, \Sigma(\mathbf{p},\omega_{n}) 
\nonumber \\
& \simeq & - \int \! \frac{d\mathbf{p}}{(2 \pi)^{3}} \frac{1}{\beta} \sum_{n} \mathcal{G}_{0}(\mathbf{p},\omega_{n})^{2}\,  \mathcal{G}_{0}(-\mathbf{p},-\omega_{n})
\nonumber \\
& \times & \int \! \frac{d\mathbf{q}}{(2 \pi)^{3}} \frac{1}{\beta} \sum_{\nu} e^{i \Omega_{\nu} \eta} \, \Gamma(\mathbf{q},\Omega_{\nu})
\label{fermionic-density-uno}
\end{eqnarray}
owing to the approximate form for the self-energy (\ref{t-matrix-self-energy}) which is valid in this limit.
With the polar approximation (\ref{polar-particle-particel-propagator}) for $\Gamma(\mathbf{q},\Omega_{\nu})$ and the further approximate result (cf., e.g., Section 3.1 of Ref.~\cite{Physics-Reports-2018})
\begin{equation}
 \int \! \frac{d\mathbf{p}}{(2 \pi)^{3}} \frac{1}{\beta} \sum_{n} \mathcal{G}_{0}(\mathbf{p},\omega_{n})^{2}\,  \mathcal{G}_{0}(-\mathbf{p},-\omega_{n}) \simeq - \, \frac{m^{2} \, a_{F}}{8 \pi} \, ,
\label{further-approximation}
\end{equation}
the expression (\ref{fermionic-density-uno}) reduces to 
\begin{equation}
n_{\sigma}^{(1)} = \int \! \frac{d\mathbf{q}}{(2 \pi)^{3}} \, \frac{1}{e^{\beta \xi_{B}(\mathbf{q})} - 1}
\label{fermionic-density-uno-final}
\end{equation}
which coincides with the density $n_{\mathrm{p}}$ of bosons (molecules) given by Eq.(\ref{approximate-bosonic-density}).
A combination of Eqs.~(\ref{approximate-fermionic-density-equation}), (\ref{fermionic-density-zero}), and (\ref{fermionic-density-uno-final}) yields eventually the result:
\begin{eqnarray}
n_{\sigma} = n_{\mathrm{f}} + n_{\mathrm{p}} & = & \int \! \frac{d\mathbf{p}}{(2 \pi)^{3}} \, \frac{1}{e^{\beta \xi(\mathbf{p})} + 1}
\nonumber \\
& + & \int \! \frac{d\mathbf{q}}{(2 \pi)^{3}} \, \frac{1}{e^{\beta \xi_{B}(\mathbf{q})} - 1} \, .
\label{density-decomposition}
\end{eqnarray}
At this point, the fermionic chemical potential $\mu$ can be eliminated from Eq.~(\ref{density-decomposition}) by fixing the value of $n_{\sigma}$ therein, with the bosonic chemical potential 
$\mu_{B} = 2 \mu + \varepsilon_{0}$ following in a consistent way.

There remains to find an explicit connection with the expressions of the fermion-boson (atom-molecule) model, which were obtained in Refs.~\cite{Chin-Grimm-2004,Eagles-1969} in the classical limit and used in  Ref.~\cite{Ulm-Cam-2019} to account for the experimental data in the BEC regime of the phase diagram.
To this end, we consider the classical limit of the expressions (\ref{density-decomposition}) by neglecting $\pm 1$ in the denominators, and perform the trap average by replacing 
$\mu \rightarrow \mu - V_{\mathrm{f}}(\mathbf{r})$ and $\mu_{B} \rightarrow \mu_{B} - V_{\mathrm{p}}(\mathbf{r})$ and integrating over the space variable $\mathbf{r}$, similarly to what was done in Section \ref{sec:comparison_with_experiment}-A.
Here, 
\begin{equation}
V_{\mathrm{f}/\mathrm{p}}(\mathbf{r}) = \frac{1}{2} M_{\mathrm{f}/\mathrm{p}} \left( \omega_{x}^{2} x^{2} + \omega_{y}^{2} y^{2} + \omega_{z}^{2} z^{2} \right)
\label{anisotropic-harmonic-potential-fermion-boson}
\end{equation}
is the (anisotropic) harmonic oscillator potential commonly considered for ultra-cold gases, with $M_{\mathrm{f}} = m$ for fermions (atoms) and $M_{\mathrm{p}} = 2m$  for bosons (molecules).
The results for the total number of fermions $N_{\mathrm{f}}$ and the total number of bosons $N_{\mathrm{p}}$ then become:
\begin{equation}
N_{\mathrm{f}} \simeq \! \int \!\! d\mathbf{r} \!\! \int \! \frac{d\mathbf{p}}{(2 \pi)^{3}} e^{-\beta \left[\frac{\mathbf{p}^{2}}{2m} + V_{\mathrm{f}}(\mathbf{r}) - \mu \right]} 
= \left( \! \frac{k_{B}T}{\omega_{0}} \! \right)^{3} \! e^{\mu/k_{B}T}
\label{trap-everaged-N_f}
\end{equation}
and
\begin{equation}
N_{\mathrm{p}} \simeq \! \int \!\! d\mathbf{r} \!\! \int \! \frac{d\mathbf{q}}{(2 \pi)^{3}} e^{-\beta \left[\frac{\mathbf{q}^{2}}{4m} + V_{\mathrm{p}}(\mathbf{r}) - \mu_{B} \right]} 
= \left( \! \frac{k_{B}T}{\omega_{0}} \! \right)^{3} \! e^{\mu_{B}/k_{B}T}
\label{trap-everaged-N_p}
\end{equation}
where $\omega_{0} = (\omega_{x} \omega_{y} \omega_{z})^{1/3}$ is the average trap frequency (cf., e.g., Refs.\cite{Pitaevskii_Stringari-2003,Pethick-Smith-2008}).
From these results it follows that
\begin{equation}
\frac{N_{\mathrm{f}}^{2}}{N_{\mathrm{p}}} = \left( \! \frac{k_{B}T}{\omega_{0}} \! \right)^{3} e^{(2\mu - \mu_{B})/k_{B}T} = \left( \! \frac{k_{B}T}{\omega_{0}} \! \right)^{3} e^{-\varepsilon_{0}/k_{B}T} \, ,
\label{N_f-square-over-N_p-trapped}
\end{equation}
from which, by replacing $\omega_{0} = E_{F} / (6N_{\sigma})^{1/3}$ where $E_{F}$ is the Fermi energy for the trap, one recovers the expression reported in Appendix A of
Ref.~\cite{Ulm-Cam-2019}.
More generally, $N_{\mathrm{p}}$ and $N_{\mathrm{f}}$ for the trapped case could be obtained in closed form directly from Eqs.~(\ref{approximate-bosonic-density}) and (\ref{fermionic-density-zero}), in terms of $\text{Li}_3(e^{\beta\mu_{B}})$ for bosons and $\text{Li}_3(-e^{\beta\mu_{B}})$ for fermions (where $\text{Li}_n(z)$ is the poly-logarithmic function of index $n$ and argument $z$).
The expression (\ref{N_f-square-over-N_p-trapped}) generalizes to a harmonically trapped system the \emph{law of mass action} valid for a homogeneous system \cite{LL-1999}.

Finally, it is worth summarizing what is lost when passing from the fully quantum many-body approach to its simplified version obtained above.
To get this simplified version, in Eq.~(\ref{G_B-FT-sc}) we have 
(i) neglected the ``unbound'' term $-\mathcal{F}_{2}(\mathbf{q},\Omega_{\nu})$, 
(ii) approximated $\phi({\mathbf p})$ in the expression (\ref{def-F-j}) for $\mathcal{F}_{1}(\mathbf{q},\Omega_{\nu})$ by the two-body form (\ref{psi-k-BEC}) and taken $\mu = - \varepsilon_{0}/2$ therein
with $\varepsilon_{0} \gg k_{B} T$,
and (iii) approximated $\Gamma(\mathbf{q},\Omega_{\nu})$ by the polar form (\ref{polar-particle-particel-propagator});
while in Eq.~(\ref{fermionic-density-equation}) we have performed the expansion (\ref{G-expanded}) with the typical approximations that apply to the BEC limit at low temperature when $\mu$ is the largest energy scale
in the problem.
None of these approximations, however, is valid either away from the BEC limit when approaching unitarity at any temperature, or in the BEC limit itself for sufficiently high temperature.
In both these cases, the fermionic nature of the ``preformed pairs'' manifests itself and only fermionic correlations remain physically relevant.
On physical grounds, the results of the quantum many-body approach and of the statistical fermion-boson model differ from each other to the extent that the latter bears essentially on the chemical 
reaction (dimer $\longleftrightarrow$ spin-$\uparrow$ + spin-$\downarrow$) for
molecules that break up into atom pairs and vice-versa, with no regard on the way the molecules are formed by the laws of quantum mechanics and on the effects that the surrounding 
environment might exert on them through inter-particle collisions.

\begin{figure}[t]
\begin{center}
\includegraphics[width=8.5cm,angle=0]{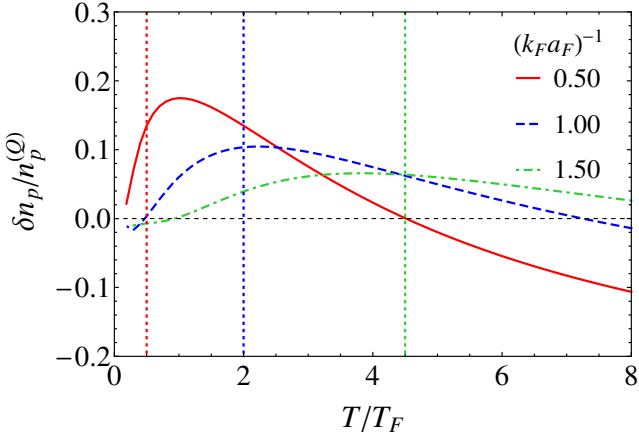}
\caption{(Color online) Temperature dependence of the relative difference $\delta n_{\mathrm{p}}/n_{\mathrm{p}}^{(Q)}$ with $\delta n_{\mathrm{p}} = n_{\mathrm{p}}^{(Q)} - n_{\mathrm{p}}^{(C)}$ 
                                     between the quantum many-body (Q) and classical statistical (C) calculations of the pair density $n_{\mathrm{p}}$ for the homogeneous system at various couplings. The vertical lines indicate the corresponding binding energies $\varepsilon_0$ (in units of $E_F$), for increasing coupling from left to right.}
\label{Figure-11}
\end{center}
\end{figure}

In this context, it is interesting to explicitly verify to what extent the results of the quantum many-body approach (Q) and of the classical statistical model (C) differ from each other in the BEC limit of
the homogeneous system at sufficiently high temperature.
To this end, Fig.~\ref{Figure-11} shows the temperature dependence of the relative difference $\delta n_{\mathrm{p}}/n_{\mathrm{p}}^{(Q)}$ 
for the couplings $(k_{F} a_{F})^{-1} = (0.5,1.0,1.5)$, where $\delta n_{\mathrm{p}} = n_{\mathrm{p}}^{(Q)} - n_{\mathrm{p}}^{(C)}$. 
One sees that this relative difference can be substantial in all cases.
In particular, for $k_{B} T \lesssim \varepsilon_{0}$ the relative difference increases with increasing temperature and decreases with increasing coupling, as expected.
The following apparent reduction of the relative difference for $k_{B} T \gtrsim \varepsilon_{0}$ then turns into a substantial increase (in absolute value) when $k_{B} T \gg \varepsilon_{0}$. 
Again in favor of the results obtained by the quantum many-body ($t$-matrix) approach, one should recall that in the high-temperature limit this approach correctly recovers the controlled high-temperature (virial) expansion to second order \cite{Combescot-2006}. 
Specifically, when this high-temperature expansion is made on the self-energy, keeping both the bound-state (pole) and scattering (continuum) contributions to the particle-particle propagator $\Gamma$
of Eq.~(\ref{Gamma}) turns out to be essential to correctly recover the virial expansion.  
Since the statistical model includes only the bound-state contribution, it unavoidably fails in the high-temperature limit.

\section{CRITICAL TEMPERATURE OF A LOW-DENSITY TRAPPED BOSE GAS}
\label{sec:appendix-scattering-length}

In this Appendix, we calculate the superfluid critical temperature of a low-density Bose gas in a trap, where the interaction is treated at the level of the two-body $t$-matrix specified by the scattering length $a_{B}$.
Similarly to what we did in Section \ref{sec:comparison_with_experiment}-A for the trapped Fermi gas, we adopt a local-density approach whereby the bosonic chemical potential $\mu_{B}$ is replaced by a 
local chemical potential $\mu_{B}(\mathbf{r})$.
We thus write for the bosonic density
\begin{equation}
n_{B}(\mathbf{r}) = \int \! \frac{d\mathbf{q}}{(2 \pi)^{3}} \, \frac{1}{e^{\beta \left[\frac{\mathbf{q}^{2}}{2m_{B}} - \mu_{B}(\mathbf{r}) \right]} - 1}
\label{bosonic-density-C}
\end{equation} 
where $\mu_{B}(\mathbf{r}) = \mu_{B} - V_{B}(\mathbf{r}) - 2 t_{0}n_{B}(\mathbf{r})$.
Here, $V_{B}(\mathbf{r})$ is the trapping potential of the form (\ref{anisotropic-harmonic-potential-fermions}) with $m \rightarrow m_{B}$ (we also assume 
$\omega_{x}=\omega_{y}=\omega_{z}=\omega_{0}$ for simplicity), and $2 \, t_{0} \, n_{B}(\mathbf{r})$ is the leading approximation to the
self-energy of a dilute Bose gas in the normal phase where $t_{0} = 4 \pi a_{B} / m_{B}$ \cite{Popov-1987}.
Note that, owing to the presence of the local self-energy $2 \, t_{0} \, n_{B}(\mathbf{r})$, Eq.~(\ref{bosonic-density-C}) is a self-consistent condition for $n_{B}(\mathbf{r})$.
Once $n_{B}(\mathbf{r})$ is known, the total number of bosons is obtained as follows:
\begin{equation}
N_{B} = \int \! d\mathbf{r} \, n_{B}(\mathbf{r}) \, .
\label{total-number-boson-C}
\end{equation}

We are interested in determining the dependence on $N_{B}$ of the critical temperature $T_{c}$ for the transition to the superfluid phase.
Similarly to what happens for a trapped Fermi gas (cf.~Section \ref{sec:comparison_with_experiment}-A), also for a trapped Bose gas the central portion of the cloud density is where superfluidity first manifests itself upon lowering the temperature from the normal phase.
At $\mathbf{r} = 0$, the Hugenholtz-Pines condition \cite{HP-1959} for $T_{c}$ then yields
\begin{equation}
\mu_{B} = 2 \, t_{0} \, n_{B}(\mathbf{r}=0) 
\label{Hugenholtz-Pines-condition-C}
\end{equation}
for the thermodynamic bosonic potential in the trap.
At $T_{c}$, we can then write $\mu_{B}(\mathbf{r}) = - V_{B}(\mathbf{r}) - 2 \, t_{0} \, \delta n_{B}(\mathbf{r})$ with $\delta n_{B}(\mathbf{r}) = \left[ n_{B}(\mathbf{r}) - n_{B}(\mathbf{r}=0) \right]$, such that
Eq.~(\ref{bosonic-density-C}) becomes:
\begin{equation}
n_{B}(\mathbf{r}) = \int \! \frac{d\mathbf{q}}{(2 \pi)^{3}} \, \frac{1}{e^{\beta_{c} \left[\frac{\mathbf{q}^{2}}{2m_{B}} + V_{B}(\mathbf{r}) + 2 t_{0} \delta n_{B}(\mathbf{r}) \right]} - 1}
\label{bosonic-density-at-Tc-C}
\end{equation}
where $\beta_{c} =(k_{B} T_{c})^{-1}$.
For any given value of $\mathbf{r}$, this equation is solved self-consistently for the variable $n_{B}(\mathbf{r})$ by fixing an arbitrary value of $n_{B}(\mathbf{r}=0)$ to start with, in such a way that $n_{B}(\mathbf{r})$ 
never exceeds $n_{B}(\mathbf{r}=0)$.
Once the entire density profile $n_{B}(\mathbf{r})$ is obtained in this way, one calculates $N_{B}$ from Eq.~(\ref{total-number-boson-C}) so as to obtain $T_{c}$ as a function of $N_{B}$ and $a_{B}$.
In addition, upon measuring the values of $T_{c}$ obtained in this way in units of the critical temperature for non-interacting trapped bosons $k_{B} T_{c}^{\mathrm{BEC}} = \omega_{0} \left[ N_{B}/\zeta(3)\right]^{1/3}$ 
(where $\zeta(z)$ is the Riemann zeta function of argument $z$), one finds that $T_{c}/T_{c}^{\mathrm{BEC}}$ is a function only of the scaling variable $a_{B} \sqrt{k_{B} T_{c}^{\mathrm{BEC}}/m_{B}}$.
By translating back into the language of the BCS-BEC crossover of the main text, one gets eventually that $T_{c}/T_{F}$ is a function of the coupling parameter $(k_{F} a_{F})^{-1}$ in the trap since $a_{B}$ is proportional to $a_{F}$ 
(cf.~Fig.~\ref{Figure-7}). 



\end{document}